\documentclass[10pt,letterpaper,twocolumn]{article}

\usepackage[letterpaper,top=1in,bottom=1in,left=0.75in,right=0.75in,columnsep=0.3in]{geometry}
\usepackage[T1]{fontenc}
\usepackage{times}
\usepackage{amsmath,amssymb,amsfonts}
\usepackage{graphicx}
\usepackage{textcomp}
\usepackage{xcolor}
\usepackage{booktabs}
\usepackage{array}
\usepackage{tabularx}
\usepackage[hyphens]{url}
\usepackage{hyperref} 
\newcommand{\sep}{\ $\cdot$\ }

\newcolumntype{L}{>{\raggedright\arraybackslash}X}
\newcolumntype{s}[1]{>{\hsize=#1\hsize\raggedright\arraybackslash}X}

\usepackage{stfloats}
\let\CAMEOoldbibliography\thebibliography
\renewcommand{\thebibliography}[1]{%
  \CAMEOoldbibliography{#1}%
  \setlength{\itemsep}{0pt}%
  \setlength{\parsep}{0pt}%
  \setlength{\parskip}{0pt}}

\usepackage{tikz}
\usetikzlibrary{shapes.geometric, arrows.meta, positioning}

\graphicspath{{figures/}}

\begin{document}

% ---------------------------------------------------------------
% Title block, author, abstract and keywords (spanning both columns)
\twocolumn[
\begin{center}
{\LARGE\bfseries CAMEO: A Class-Activation-Mapped Equitable Overlay Framework for
Fair and Robust Deep Learning-based Skin Condition Diagnosis\par}
\vspace{1.2em}
{\large Youssef Attia and Debasmita Mukherjee*\par}
\vspace{0.6em}
{\small\itshape Department of Electrical and Computer
Engineering, University of New Brunswick, Fredericton, NB, Canada\par}
\vspace{1.2em}
\begin{minipage}{0.88\textwidth}
\small
\begin{center}\textbf{Abstract}\end{center}
\vspace{-0.5em}
Deep learning classifiers for dermoscopic skin lesions often reach high
in-distribution accuracy while quietly relying on spurious background
cues such as skin tone, device vignetting and embedded rulers, rather
than on lesion morphology. This undermines robustness and fairness
across skin tones. This work asks whether Explainable AI (XAI),
typically used only to audit a finished model, can instead be
repurposed as an \emph{active training signal} that corrects this
shortcut without sacrificing diagnostic accuracy. We introduce
\textbf{CAMEO} (Class-Activation-Mapped Equitable Overlay), a framework
that answers this question affirmatively. \textbf{CAMEO} improves skin-lesion classification by selecting stable model explanations and using them to separate lesions from their backgrounds.
It then replaces the background with realistic synthetic skin while keeping the lesion unchanged.
On HAM10000 and dark-skin ISIC images, CAMEO maintained accuracy while reducing background-driven errors by nearly four times.
It also made the model’s attention more consistent when backgrounds changed.
Results across multiple tests show that reducing reliance on background information improves robustness, with Fitzpatrick-based backgrounds providing a realistic and interpretable approach.
Taken
together, the results show that XAI-guided augmentation can make
dermoscopic classifiers measurably more robust and fair at no cost to
accuracy. They also clarify that it is the \emph{mechanism} and not the
specific tone palette that matters, and that the lasting contribution
of XAI here lies in stability-screened, annotation-free lesion
localisation rather than in the robustness number itself.

\end{minipage}
\end{center}
\vspace{1em}
]

% ---------------------------------------------------------------
\section{Introduction}

\subsection{Motivation: access, trust, and skin tone}

Long waits for diagnosis and treatment are common in public health
systems. In Canada, the median wait from general-practitioner referral
to treatment was 28.6 weeks in 2025~\cite{fraser2025}, against a global
average of 70 days~\cite{philips2025}. Automated screening could help,
and deep learning already detects skin cancer from images at a level
useful for triage~\cite{esteva2017}. Uptake is still limited, mainly
because of trust. Explainable Artificial Intelligence (XAI) is a popular area of research
~\cite{wilkinson2026}, but explanations have not closed the trust gap:
61\% of healthcare professionals in the Future Health Index survey named
data bias as a barrier to trsusting and adopting AI~\cite{philips2025}, and bias in
the data is inherited by the models trained on it.

Dermatology demonstrates this bias clearly. Skin tone is described clinically by the
Fitzpatrick scale, six ordinal levels from type~I (very light skin that
always burns and never tans) to type~VI (deeply pigmented brown or black
skin that never burns)~\cite{fitzpatrick1988}; it is the standard record
of constitutive pigmentation and the scale we use throughout. Erythema
(skin redness) is a main visual sign of early pressure injury, infection
and burns, but on Fitzpatrick types V and VI it often appears purple,
brown or grey rather than red~\cite{dhoonmoon2021,aoki2026}. Most
wound-care and skin-cancer datasets contain lighter skin, so models miss
early disease in darker-skinned patients~\cite{aoki2026}, and because
redness is harder to see, inflammatory diseases such as atopic
dermatitis are underdiagnosed or scored as milder on darker
skin~\cite{shao2022,mckenzie2022,forsyth2025}. Atopic dermatitis is also
more common, and more often severe, in Black and Hispanic
patients~\cite{kaufman2018,bengashir2002}. Fair diagnosis therefore
needs AI methods that reduce racial bias in the data.

\subsection{From skin tone bias to shortcut learning}

The classifiers used for this task are supervised deep convolutional
neural networks (CNNs), typically an ImageNet-pretrained backbone
fine-tuned to predict a diagnostic label from a single dermoscopic
image. They make the same mistake at the level of pixels: trained on
standard benchmarks, they routinely learn spurious background
correlations instead of the lesion morphology that carries the
diagnostic information~\cite{esteva2017,kinyanjui2020,bissoto2019}.
Three such correlations recur in dermoscopy datasets:

\begin{itemize}
    \item \textbf{Background skin tone.} The healthy skin around the
    lesion carries no diagnostic information, but if one diagnosis is
    photographed more often on lighter skin, the model can read that
    surrounding tone as evidence for the label.
    \item \textbf{Device vignetting.} Dermatoscopes darken the image
    corners, because the lens and contact plate restrict light reaching
    the sensor edge. The shape of that dark ring differs between devices
    and therefore between contributing clinics, so it acts as a hidden
    site label.
    \item \textbf{Embedded measurement rulers.} Clinicians place a
    graduated adhesive ruler beside a lesion to record its size, and
    they measure the lesions that worry them, so a ruler correlates with
    malignancy for reasons unrelated to how the lesion looks.
\end{itemize}

These shortcuts are not evident in balanced in-distribution evaluation but
appear as performance gaps across skin tones, devices and clinical
settings~\cite{daneshjou2022,groh2021}. We evaluate on melanoma because
that task has public labelled data and published fairness baselines, but
the mechanism we propose edits the skin surrounding a lesion and leaves
the lesion untouched, so it transfers to other conditions whose
appearance changes with pigmentation (Section~\ref{sec:limits}). The
primary research question of this work is therefore:

\begin{quote}
\textit{Can XAI-guided, semantically-aware data augmentation improve the
robustness and cross-skin-tone fairness of deep learning models for
dermoscopic skin lesion classification, without sacrificing diagnostic
accuracy?}
\end{quote}

\subsection{Overview of the proposed framework}
\label{ssec:overview}

Our proposed solution is \textbf{CAMEO} (Class-Activation-Mapped Equitable
Overlay), and the idea is simple to state: ask the classifier which
pixels it is using, check that the answer is trustworthy, then repaint
only the pixels that should not have mattered.

Two terms are needed to follow the paper.An \emph{attribution map}, also called a \emph{saliency 
map}~\cite{simonyan2014}, is an image the size of the input 
in which each pixel is scored by how much it contributed to 
the prediction, so bright regions are the evidence the model 
used. It is produced from
the model, and therefore reports what that particular model looked at
rather than what a clinician would. We call such a map \emph{stable}
when recomputing it after a change to the input too small for a human to
see leaves the highlighted region essentially unchanged, and
\emph{unstable} otherwise.

CAMEO runs in four steps. It computes attribution maps for every
training image and screens them for stability, keeping the score as a
per-image confidence value. It suppresses the measurement-ruler artifact
that otherwise dominates those maps and fuses the corrected maps into a
binary lesion mask, using no segmentation labels. It then applies
\textbf{ChromaSwap}, the background-replacement augmentation we
introduce here, which holds the lesion interior fixed and repaints the
surrounding skin with a textured synthetic Fitzpatrick tone. Finally it
retrains the classifier and evaluates it, separating genuine errors from
cases that simply fall outside the training distribution.

Two of these decisions carry the novelty of the framework and we return
to them in Section~\ref{sec:gap}: the stability screen, which decides
which explanations are safe to act on before any pixel is edited, and
the choice to repaint with calibrated skin tones rather than arbitrary
textures.

% ---------------------------------------------------------------
\section{Related Work}

\textbf{Fairness in dermatology AI.}
Kinyanjui~et~al.~\cite{kinyanjui2020} documented accuracy disparities
across Fitzpatrick types in dermatology classifiers, establishing the
fairness motivation for this work, and Groh et~al.~\cite{groh2021} and
Daneshjou et~al.~\cite{daneshjou2022} further quantified degraded
performance on darker skin. Yang et~al.~\cite{yang2023} proposed
adversarial debiasing, which trains a second network to stop a predictor
leaking sensitive attributes; it is effective on tabular electronic
health record data, but needs explicit sensitive-attribute labels and
operates on feature vectors rather than the spatial image structure
where skin-tone bias resides. Our approach needs no demographic labels:
Fitzpatrick diversity is injected directly into the image through
background replacement, and the region to replace is detected without
supervision. This is more than a data-preparation step, because the
region is chosen by the model being corrected, the correction strength
is set by how trustworthy that model's own explanation is, and the
corrected images are fed back into training. CAMEO is therefore a closed
loop between explanation and optimisation, not a fixed filter.

\textbf{Lesion segmentation.}
SkinSAM~\cite{skinsam2023} fine-tunes the Segment Anything Model on
HAM10000 ground-truth masks and reaches mean IoU\,=\,0.785 with
bounding-box prompts, but without prompting its reported mean IoU falls
to about 0.16, so its strong numbers presuppose a localisation hint
unavailable in the annotation-free setting we target.
DermNet~\cite{dermnet2026} applies SAM with colour-space thresholding as
a pre-processing step. Our XAI-derived mask is less precise on boundary
metrics than a fine-tuned SAM, reaching IoU\,=\,0.422 against 0.785, a
gap of 36.3\,pp (Section~\ref{ssec:xaigt}). It is, however, generated
automatically from the classifier's own attributions, needs no
annotation and no foundation model, is used only at training time, and
beats the same SAM model by about $2.6\times$ when that model is denied
the prompt our setting cannot supply.

\textbf{Explainability methods.}
We build on GradCAM++~\cite{chattopadhay2018}, which generalises
gradient-weighted class-activation mapping~\cite{selvaraju2017}, and on
Integrated Gradients~\cite{sundararajan2017}. Explanation stability is
measured with the Structural Similarity Index~\cite{wang2004}, and
adversarial brittleness with the Fast Gradient Sign Method
(FGSM)~\cite{goodfellow2015}. Recent surveys of trustworthy
medical-imaging AI~\cite{salahuddin2025}, of XAI across machine learning
paradigms~\cite{wilkinson2026} and of robust, fair foundation
models~\cite{chen2024} all argue that XAI must move from a reporting
tool to an active component of the training loop, which is our aim.
Those surveys also note that attribution maps are model-dependent and
seldom validated before use, motivating the stability screen of
Section~\ref{ssec:stability}.

\textbf{Bias-adversarial augmentation.}
BiasAdv~\cite{biasadv2023} generates bias-conflicting synthetic samples
by adversarially attacking an auxiliary biased model. We differ in
mechanism, using XAI attribution maps rather than adversarial
perturbations; in domain, working on dermoscopy rather than natural
images; and in interpretability, since ChromaSwap images can be
inspected visually whereas adversarial perturbations are imperceptible
by construction.

\textbf{Domain adaptation and covariate shift.}
Our previous works ~\cite{mukherjee2025, Mukherjee_ccece, Mukherjee_2023, Mukherjee_pers} introduced SIMLea, which
uses feature-space statistical similarity to quantify covariate shift
and decide when transfer learning is feasible.We adapt its $R^{2}$ feature-space distance to split our 
dark-skin external test set into in-distribution and 
out-of-distribution subsets. Concretely, $R^{2}$ here 
measures how well a new image's CNN feature vector can be 
reconstructed from the principal directions learned on the 
training data: a score near~1 means the image is 
statistically familiar to the model, while a score well 
below~1 signals genuine novelty. An image is labelled 
in-distribution when its score exceeds a threshold set at 
the 5th percentile of in-domain validation scores, following 
the out-of-distribution detection baseline of Hendrycks and 
Gimpel~\cite{hendrycks2017}. The full mathematical 
formulation, including the PCA basis fitting and the 
per-feature shift scalar $R_{\mathrm{shift}}$, is given in 
Section~\ref{ssec:r2}.Both terms recur
below: an image is \emph{in-distribution} when its features resemble
those seen in training and \emph{out-of-distribution} when they do not.
The distinction matters for fairness, because a model failing on
out-of-distribution images is being asked a question its training data
never contained.

% ---------------------------------------------------------------
\section{Research Gap and Contributions}
\label{sec:gap}

Having reviewed the literature, we can now state precisely what is
missing and what this paper adds.

\subsection{Research Gap}

Current skin-condition classifiers share five connected limitations.

\begin{itemize}
    \item \textbf{Shortcut reliance.} Models trained on public
    dermoscopy benchmarks exploit background skin tone, device
    vignetting and embedded rulers as predictive
    cues~\cite{kinyanjui2020,bissoto2019}. Bissoto
    et~al.~\cite{bissoto2019} showed the extreme case: a classifier
    trained on images with the lesion erased still beat a
    dermatologist-calibrated benchmark, so its signal lay entirely
    outside the lesion.
    \item \textbf{Spatially uniform augmentation.} Standard pipelines
    transform the whole image, leaving the model free to re-learn
    background shortcuts from the unperturbed
    portion~\cite{shorten2019}. Nothing in a rotation or colour jitter
    tells the model which part of the image it may use.
    \item \textbf{No artifact awareness.} Imaging artifacts such as the
    measurement ruler are high-contrast structures that attract both the
    classifier and the attribution method, so any pipeline deriving a
    lesion region from attributions inherits the artifact unless it
    removes it deliberately. We are not aware of a dermoscopy
    augmentation method that suppresses known artifacts inside the loop
    generating its own supervision signal.
    \item \textbf{Post-hoc and unvalidated XAI.} Explainability tools
    are used almost exclusively to interpret a finished model rather
    than correct it~\cite{salahuddin2025}, which is a problem twice
    over. An explanation produced after training can document a bias but
    not remove it, so audit and fix stay separate and the fix usually
    needs new labelled data. More seriously, on the rare occasions when
    attributions do alter training, their stability is rarely checked
    first, even though attributions are model-dependent and least
    reliable on exactly the images where the model is already wrong.
    \item \textbf{Demographic generalisation gap.} Most skin-condition
    datasets are heavily skewed toward light skin, Fitzpatrick~I to III,
    and models trained on them are not systematically evaluated on, or
    corrected for, darker skin tones~\cite{daneshjou2022,groh2021}.
\end{itemize}

\subsection{Novelty}

CAMEO addresses all five above mentioned research gaps within one self-contained pipeline, and two
design decisions separate it from existing work.

\textbf{First, CAMEO decides which explanations are safe to use before
using them.} An attribution map comes from the model, so it inherits
whatever the model has learned, including its biases. If a change to the
input too small to see moves the highlighted region, the resulting mask
describes a fragile shortcut rather than the lesion, and repainting on
that basis would write a new model-specific artifact into the training
data, reinforcing the very bias we set out to remove. CAMEO therefore
recomputes every attribution map under small Gaussian perturbations and
scores the agreement with the Structural Similarity Index
(SSIM)~\cite{wang2004}; that score decides how the sample is used
downstream (Section~\ref{ssec:stability}). Anyone can threshold a
saliency map and repaint a background, but without this screen nothing
guarantees the edit is anchored to the lesion. We regard the screen, not
the repainting, as what makes XAI-guided augmentation safe inside a
training loop, and to our knowledge no prior dermatology debiasing
method gates its augmentation on measured explanation stability.

\textbf{Second, CAMEO replaces the background with visually identifiable features the model
still recognises as skin.} Our ChromaSwap augmentation draws replacement
tones from the Fitzpatrick I to VI scale and textures them rather than
filling flat, so an augmented image still looks like a dermoscopic image
of human skin. An arbitrary texture would push the sample away from the
deployment distribution and turn a useful counterfactual into an
anomaly. Calibrated replacement keeps every augmented sample
in-distribution or within a manageable shift, which also lets the
$R^{2}$ feature-space test we use on external data
(Section~\ref{ssec:r2}) be applied to the augmented data itself. The
cost of ignoring this is evident through our ablations, where a flat,
unrealistic background destroys robustness (Section~\ref{ssec:ablation}).

\subsection{Contributions}

\begin{enumerate}
    \item \textbf{An explanation-stability screen that gates the whole
    pipeline.} Every GradCAM++~\cite{chattopadhay2018} and Integrated
    Gradients~\cite{sundararajan2017} map is scored for stability under
    small input perturbations, using SSIM~\cite{wang2004}, before it may
    influence training, and that score also sets how strongly each image
    is augmented.
    \item \textbf{An XAI-guided, artifact-aware augmentation pipeline.}
    We suppress the measurement-ruler artifact in both attribution maps,
    fuse the corrected maps into a clean lesion mask, and use it to
    drive lesion-preserving background replacement with tones calibrated
    to real skin, needing no external segmentation annotation and no
    foundation model.
    \item \textbf{Debiasing through shortcut correction, benchmarked
    against established methods.} We use \emph{debiasing} throughout to
    mean removing the statistical link between background skin tone and
    the predicted label. A background-counterfactual test shows the
    melanoma decision barely moves when only the tone changes, and a
    GradCAM++ analysis shows the same for where the model looks. Across
    five seeds we compare CAMEO against a random-background control, an
    adversarial gradient-reversal baseline~\cite{yang2023,ganin2016} and
    BiasAdv~\cite{biasadv2023}: the gain comes from background
    decorrelation in general, and Fitzpatrick calibration is the
    clinically realistic instance of that mechanism, matching or
    exceeding every baseline without demographic labels.
    \item \textbf{Annotation-free, model-faithful lesion localisation.}
    Our composite XAI masks reach mean Intersection over Union
    (IoU)\,=\,0.422 against 2{,}332 dermatologist-drawn segmentations,
    against IoU\,$\approx$\,0.16 for the Segment Anything Model (SAM)
    without a prompt~\cite{skinsam2023}. They need no intensity
    heuristic, manual annotation or external model, and because they are
    derived from the classifier being corrected, they show what it
    actually uses rather than what an independent segmenter would
    outline.
    \item \textbf{Covariate-shift-aware fairness evaluation.} An $R^{2}$
    feature-space filter with a percentile out-of-distribution threshold
    separates genuine classification error from domain mismatch on an
    external dark-skin test set, so a model is not penalised for data
    statistically foreign to its training distribution.
\end{enumerate}

% ---------------------------------------------------------------
\section{Methodology}

\subsection{Datasets and Notation}
\label{ssec:data}

Training and in-distribution testing use \textbf{HAM10000} (Human
Against Machine with 10{,}000 training images)~\cite{tschandl2018}, a
public archive of 10{,}015 dermatoscopic images of common pigmented skin
lesions, collected over twenty years at a clinic in Austria and a
practice in Australia and labelled across seven diagnostic categories.
It is the most widely used public benchmark for skin-lesion
classification, which is why we adopt it, and it is strongly skewed
toward light skin, which makes it a suitable subject for this study. We
reformulate the task as binary classification: melanoma versus
non-melanoma. External testing uses the \textbf{ISIC Archive}, the
public repository of the International Skin Imaging Collaboration, which
aggregates dermoscopic images from many centres and hosts the annual
ISIC challenges~\cite{codella2019}. Drawing on more sites than HAM10000,
it contains a small number of images of darker-skinned patients, from
which we build a genuinely external dark-skin test set.

A \emph{patient-aware} stratified split of HAM10000 using
\texttt{lesion\_id} as the grouping key prevents images of the same
lesion from appearing in more than one partition (training 70\%,
validation 15\%, test 15\%). A \emph{frozen, class-balanced} test set of
342 images (171 melanoma and 171 non-melanoma) is held fixed across all
experiments and never used for model selection or hyper-parameter
search.

Because the three evaluation sets are easy to confuse, we fix the
notation here, before any of it is used (Table~\ref{tab:notation}).
\textbf{T0} is the frozen class-balanced HAM10000 test set just
described, in-distribution by construction. \textbf{T1} is the
\emph{external} dark-skin generalisation set, 100 ISIC images (50
melanoma and 50 non-melanoma, all HAM10000 identifiers excluded),
measuring real cross-tone shift. \textbf{T\_swap} is a \emph{synthetic}
background-counterfactual set in which each T0 lesion is re-composited
onto a replacement skin-tone background, measuring controlled background
shift with everything else fixed. T1 and T\_swap probe different
questions and are never compared with each other, and every accuracy we
report names the set it was measured on.

\begin{table}[!htbp]
\caption{Dataset notation used throughout. T1 (external, real) and
T\_swap (synthetic, controlled) are different sets and are never
compared directly. Two probes use T\_swap-style compositing: the
decision-flip and attention probe (Section~\ref{ssec:swap}) re-composites
150 lesions onto six tones, whereas the T\_swap accuracy metric uses one
fixed swap per T0 image ($N{=}342$).}
\centering
\scriptsize
\setlength{\tabcolsep}{4pt}
\begin{tabularx}{\columnwidth}{@{}l c L@{}}
\toprule
\textbf{Set} & \textbf{$N$} & \textbf{Description and purpose} \\
\midrule
T0        & 342 & Frozen class-balanced HAM10000 test set (in-distribution). \\
T1        & 100 & External dark-skin ISIC set (real cross-tone shift). \\
T\_swap   & 342 & Synthetic background swap of T0 (controlled robustness). \\
\bottomrule
\end{tabularx}
\label{tab:notation}
\end{table}

\subsection{Baseline Models}

An EfficientNet-B4 backbone~\cite{tan2019} pretrained on ImageNet is
used as the classifier ($380\times380$ input). Two raw-image baselines
are established before any augmentation. \textbf{Base-Skewed} is trained
on the full, naturally imbalanced training set (6{,}224 non-melanoma and
774 melanoma), and \textbf{Base-Balanced} on a class-balanced subset
(774 and 774) obtained by undersampling the majority class;
Base-Balanced is the primary baseline to beat.

All models share the same training recipe: AdamW optimiser,
class-weighted cross-entropy with label smoothing (0.1), cosine
learning-rate schedule with linear warm-up, mixup ($\alpha=0.2$),
gradient clipping and mixed-precision training. Early stopping uses a
\emph{balanced} validation set to avoid majority-class bias in
checkpoint selection, and hyper-parameters are fixed across all models,
so any performance difference is attributable to the data recipe rather
than extra tuning.

Table~\ref{tab:modelglossary} collects every model and baseline name in
one place, so readers can look up any name they meet later. The headline
model is called \textbf{CAMEO} throughout the text, tables and figures.
Names in the \textbf{CAMEO-} family share the same pipeline and differ
only in the training core or background strategy, whereas
\textbf{BiasAdv} and \textbf{AdvDebias-GRL} keep the names of their
original publications, being external methods rather than variants of
ours.

We developed CAMEO over four rounds, referred to as pipeline versions v1
to v4. Version~1 established the raw-image baselines; version~2 added
background replacement with a flat, untextured fill; version~3 added
texture and warm-started from the Base-Balanced weights; and version~4,
reported as CAMEO everywhere in this paper, adds the stability screen,
augments both classes symmetrically, and trains from ImageNet
initialisation rather than warm-starting. Comparing the v2 and v3
checkpoints isolates the effect of one design decision at a time, so we
keep them for the ablation in Section~\ref{ssec:ablation} under names
beginning with \textbf{Legacy-}. These are not alternative versions of
the final model. Original checkpoint filenames appear in parentheses on
first mention so results remain traceable.

\begin{table*}[!t]
\caption{Model and baseline glossary. Every name used in the paper is
defined here in one place. ``Core'' is the training data recipe and
``Background strategy'' is what replaces the region outside the lesion
mask, if anything. Identifiers from earlier pipeline versions are given
in parentheses for traceability.}
\centering
\scriptsize
\setlength{\tabcolsep}{4pt}
\begin{tabularx}{\textwidth}{@{}s{0.66} s{0.80} s{1.00} s{1.54}@{}}
\toprule
\textbf{Name} & \textbf{Core} & \textbf{Background strategy} & \textbf{Role} \\
\midrule
Base-Skewed (M1) & Full HAM10000, imbalanced (6{,}224 and 774) & None (raw images) & Naive baseline; reflects the natural class imbalance. \\
Base-Balanced (M2) & Balanced subset (774 and 774) & None (raw images) & Primary baseline to beat; class imbalance removed by undersampling. \\
\textbf{CAMEO (ours)} (M3\_AugB\_M2, v4) & Balanced subset (774 and 774) & \textbf{ChromaSwap}, symmetric on both classes & Headline model; the primary result reported throughout the paper. \\
CAMEO-Skewed (M3\_AugB\_M1) & Full HAM10000, imbalanced core & \textbf{ChromaSwap}, minority class only & Tests whether augmentation alone can substitute for explicit class balancing. \\
CAMEO-RandomBG & Balanced subset (774 and 774) & Random, non-Fitzpatrick background & Internal control: isolates whether Fitzpatrick-specific tones are necessary. \\
BiasAdv~\cite{biasadv2023} & Balanced subset (774 and 774) & Bias-conflicting background against a frozen biased classifier & External debiasing baseline, named as in its source publication. \\
AdvDebias-GRL~\cite{yang2023} & Balanced subset (774 and 774) & None; gradient reversal on border luminance & External debiasing baseline, named as in its source publication. \\
\bottomrule
\end{tabularx}
\label{tab:modelglossary}
\end{table*}

\subsection{Proposed Framework Pipeline}

\begin{figure*}[!t]
\centering
\resizebox{0.95\textwidth}{!}{
\begin{tikzpicture}[
every node/.style={font=\small},
block/.style={
    rectangle, rounded corners, draw=black, thick, align=center,
    minimum width=2.8cm, minimum height=0.95cm, fill=blue!10
},
decision/.style={
    diamond, draw=black, thick, aspect=2.2, align=center,
    fill=red!10, inner sep=1pt
},
arrow/.style={thick, ->, >=Stealth}
]
\node[block] (dataset)    at (0,0)    {Dataset\\Preprocessing};
\node[block] (baseline)   at (3.8,0)  {Baseline\\Model (Base-Balanced)};
\node[block] (xai)        at (7.6,0)  {XAI\\Extraction};
\node[decision](instab)   at (7.6,-2.4) {SSIM Stability\\Screen};
\node[block] (mask)       at (3.8,-2.4) {Ruler Fix $+$\\Composite Mask};
\node[block] (lesion)     at (0,-2.4)   {Lesion\\Extraction};
\node[block] (aug)        at (0,-4.8)   {ChromaSwap\\Augmentation};
\node[block] (retrain)    at (3.8,-4.8) {Retraining\\(ImageNet init)};
\node[block] (eval)       at (7.6,-4.8) {T0 / T1 $+$\\R$^2$ Shift Eval};
\draw[arrow] (dataset)  -- (baseline);
\draw[arrow] (baseline) -- (xai);
\draw[arrow] (xai)      -- (instab);
\draw[arrow] (instab)   -- (mask);
\draw[arrow] (mask)     -- (lesion);
\draw[arrow] (lesion)   -- (aug);
\draw[arrow] (aug)      -- (retrain);
\draw[arrow] (retrain)  -- (eval);
\end{tikzpicture}
}
\caption{Overview of the \textbf{CAMEO} pipeline. Top row: data
preparation, XAI extraction and the SSIM stability screen that decides
which explanations may be acted on and how strongly each sample is
augmented. Middle row: ruler artifact suppression and composite mask
generation. Bottom row: ChromaSwap augmentation, retraining and
evaluation. The screen is a gate, not a diagnostic: no attribution map
reaches the masking stage without passing through it.}
\label{fig:methodology_pipeline}
\end{figure*}
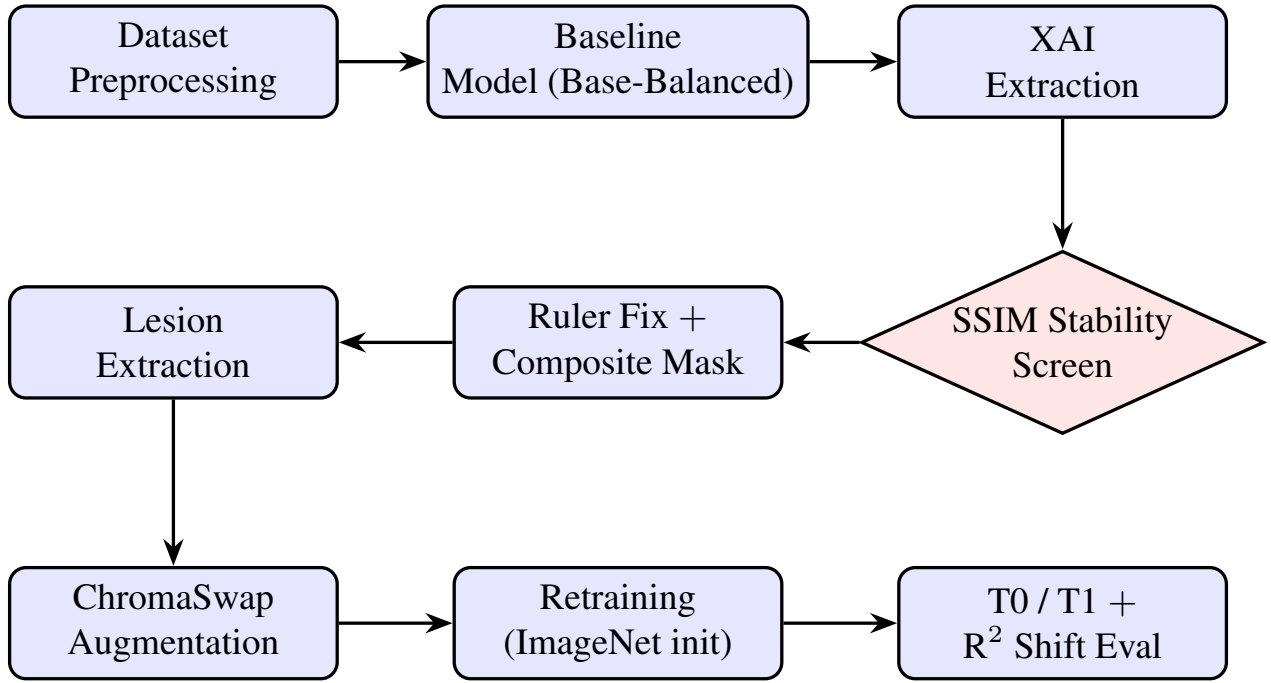

Figure~\ref{fig:methodology_pipeline} summarises the full pipeline
sketched in Section~\ref{ssec:overview}, which proceeds in four stages.

\begin{enumerate}
    \item \textbf{XAI extraction and the stability screen.} We compute
    an attribution map for every training image, then test whether it
    survives a change to the input too small for a person to notice. A
    map that survives is stable and one that moves is unstable. The
    distinction matters because the map is about to decide which pixels
    get repainted, and an unstable map reports a fragile shortcut rather
    than the lesion, so the screen sets how each image may be used.
    \item \textbf{Artifact suppression and composite mask generation.}
    We remove the measurement-ruler response from both attribution maps
    and fuse them into a single binary lesion mask.
    \item \textbf{ChromaSwap augmentation.} Our background-replacement
    augmentation uses the mask to keep the lesion interior unchanged
    pixel for pixel while repainting the surrounding skin with a
    textured synthetic Fitzpatrick tone. It is semantically aware in
    that it edits only the region the model treats as non-lesion,
    instead of transforming the image uniformly.
    \item \textbf{Retraining and evaluation.} We retrain on the
    augmented data and evaluate on T0, T1 and T\_swap
    (Table~\ref{tab:notation}).
\end{enumerate}

\subsection{Step 1: Screening Explanations for Stability}
\label{ssec:stability}

Everything downstream of this point edits training images on the basis
of what the model says it is looking at, which is only defensible if the
model says something consistent. Attribution maps are model-dependent by
construction, so a map from a shortcut-reliant classifier can point at
the background with high confidence, and a map from an uncertain
classifier can point somewhere different each time it is computed. Using
either to decide which pixels to repaint would encode the model's own
error into the dataset. We therefore treat stability screening as the
first and most important stage of CAMEO, not as a diagnostic reported
afterwards.

For each training image, GradCAM++~\cite{chattopadhay2018} and
Integrated Gradients~\cite{sundararajan2017} attributions are computed
from the Base-Balanced backbone, and an additional GradCAM pass
estimates explanation stability. Small Gaussian perturbations
($\sigma=0.01$) are applied to the input over three passes, and the
Structural Similarity Index (SSIM)~\cite{wang2004} between the original
attribution map and each perturbed map is averaged. SSIM returns 1 for
identical maps and falls toward 0 as they diverge, so the averaged value
measures how much the explanation moves when the image does not
meaningfully change. Since a perturbation of this size is imperceptible,
a high score means the map is anchored to a real image property and a
low score means it is not. Samples with mean SSIM below 0.7 are flagged
\emph{XAI-unstable}.

Flagged samples are not discarded, because medical data is scarce and
discarding a quarter of the training set would cost more than it saves.
Instead the flag \emph{modulates augmentation strength}. An XAI-unstable
image enters training as two ChromaSwap variants rather than one, each
from a different Fitzpatrick level, and its replacement background is
sampled from the full I to VI range instead of the three levels nearest
the original tone; a stable image contributes one variant from the
narrow range. The mask, feathering and texture model are identical in
both cases, so the only thing the flag changes is how much background
variation the model sees for that lesion.

The reasoning is direct. A map that moves under imperceptible noise is
tracking something other than the lesion, which is exactly where the
model most needs pushing away from background evidence, so the flag
records how much confidence that mask deserves and converts low
confidence into extra pressure rather than lost data.
Section~\ref{ssec:ssim} reports the resulting statistics.

\subsection{Step 2: Ruler Artifact Suppression and Composite Mask
Generation}
\label{ssec:mask}

A recurring artifact in HAM10000 dermoscopy is a high-intensity
rectangular measurement ruler in the upper-right corner. Because
GradCAM++ responds strongly to sharp, high-contrast edges, it
consistently highlights this diagnostically irrelevant region.

This is not a quirk of our pipeline. Artifact contamination is a
documented, general problem in skin-lesion datasets. Bissoto
et~al.~\cite{bissoto2019} found that classifiers trained on ISIC and the
Atlas of Dermoscopy stayed accurate when the lesion itself was removed,
implicating rulers, ink markings, hair and frame borders as the
surviving evidence, and Kinyanjui et~al.~\cite{kinyanjui2020} reported
the same for acquisition-correlated cues. Segmentation work on HAM10000
likewise treats hair and ruler removal as a required pre-processing
stage~\cite{dermnet2026}. What is specific to a pipeline like ours is
the consequence: because our mask is derived from the attribution map
itself, an unsuppressed artifact does not merely spoil an explanation
figure, it propagates into the mask and then into every augmented image.
Removing it is a correctness requirement, not a cosmetic one.

We suppress it before augmentation by zeroing attribution energy in the
upper-right zone (top 30\%, right 40\%) on both the GradCAM++ and the
Integrated Gradients maps \emph{before} fusion:
\[
\text{composite} = \text{norm}\!\left(\tfrac{1}{2}\bigl(
\text{rulerfix}(\text{GC++}) + \text{rulerfix}(\text{IG})\bigr)\right).
\]
The corrected maps are min-max normalised and averaged; vanilla saliency
is excluded as noisier and less focused. The composite is thresholded at
the 60th percentile and refined by a morphological clean-up chain:
outer-margin exclusion (20\%), ruler-zone exclusion, opening
($3\times3$~px), closing ($15\times15$~px),
largest-connected-component selection and dilation ($11\times11$~px).
The result is a compact binary lesion mask requiring no ground-truth
segmentation annotation.

\subsection{Step 3: Semantically-Aware ChromaSwap Augmentation}
\label{ssec:chromaswap}

Using the binary lesion mask, our \textbf{ChromaSwap} augmentation
preserves the lesion interior pixel for pixel while replacing the
background with a realistic synthetic Fitzpatrick-scale skin texture
spanning types I to VI. Both the name and the procedure are introduced
in this work. The darkest tone is a deep brown, never pure black; the
lesion boundary is Gaussian-feathered for a smooth transition; and the
background is \emph{textured} with low-amplitude noise, slight blur and
a subtle brightness gradient. Texturing is deliberate, because a flat
fill would replace one shortcut with another, constant colour being
itself an easy cue to learn. Section~\ref{ssec:fitz} shows worked
examples at every Fitzpatrick level.

\paragraph{Why not a generative model?}
An obvious alternative is to synthesise dark-skin dermoscopic images
with a generative adversarial network or a diffusion model, as has been
done for lesion synthesis in general~\cite{bissoto2018}. We chose
procedural compositing for four reasons.

\emph{Circularity.} A generative model would have to learn how lesions
look on dark skin, and the data for that does not exist. That shortage
is the problem we are working around, so the generator would inherit the
same light-skin bias as the classifier and its dark-skinned output would
be extrapolation rather than observation.

\emph{Label integrity.} A generative model can silently alter lesion
borders, pigment network and colour variegation, the exact features that
separate melanoma from a benign nevus. If the label is preserved but the
morphology is not, the sample is mislabelled in a way no automatic check
would catch. ChromaSwap cannot do this, because the lesion interior is
copied rather than generated.

\emph{Experimental control.} Our central claim is causal: the decision
changes when, and only when, the background tone changes. That needs a
counterfactual in which the background is the single difference. A
generative model changes the whole image at once, so any measured change
cannot be attributed to the background alone.

\emph{Traceability.} Every composited image can be checked by eye
against its source, the palette is a short list of published Fitzpatrick
values, and no second network must be trained, validated or reported. In
a clinical setting that matters at least as much as visual realism.

Two augmented models are trained. \textbf{CAMEO} uses the balanced real
core (774 and 774, the training data of Base-Balanced) with symmetric
ChromaSwap on \emph{both} classes ($\times1$ each), so neither class is
uniquely associated with a synthetic background. \textbf{CAMEO-Skewed}
uses the full imbalanced set (the data of Base-Skewed) with ChromaSwap
applied to the melanoma minority class to balance training (6{,}224 and
6{,}224), testing whether XAI-guided augmentation can substitute for
explicit class balancing. Both train from ImageNet initialisation with
no warm start, so any improvement over Base-Balanced is attributable to
the augmentation recipe rather than a stronger starting point.

\subsection{Step 4: Covariate-Shift Detection via $R^{2}$ Feature-Space
Similarity}
\label{ssec:r2}

Following our earlier SIMLea framework~\cite{mukherjee2025}, we quantify
covariate shift between HAM10000 (source) and any new dermoscopic set
(target) in the CNN feature space rather than in pixel space. Forward
hooks on stages 4 and 7 of the \texttt{features} module of the
Base-Balanced backbone globally average-pool activations into a vector
$\mathbf{x}\in\mathbb{R}^{D}$ per image. On a random 1{,}500-image
subset of the HAM10000 training set we fit a multivariate Gaussian and a
top-$k$ ($k=64$) PCA basis
$\mathbf{V}\in\mathbb{R}^{k\times D}$. For each target image we compute
the reconstruction $R^{2}$ in the source PCA subspace:
\[
R^{2}(\mathbf{x}) = 1 - \frac{\bigl\lVert
(\mathbf{x}-\boldsymbol{\mu}_\text{src})
-\mathbf{V}^{\!\top}\mathbf{V}
(\mathbf{x}-\boldsymbol{\mu}_\text{src})
\bigr\rVert^{2}}
{\lVert \mathbf{x}-\boldsymbol{\mu}_\text{src} \rVert^{2}}.
\]
Values close to 1 mean the image features lie in the dominant subspace
spanned by the HAM10000 training data, that is, in-distribution, while
values substantially below 1 indicate out-of-distribution novelty. The
$R^{2}$ \emph{score} is adapted from SIMLea~\cite{mukherjee2025} and the
\emph{decision rule} follows Hendrycks and Gimpel~\cite{hendrycks2017}.
We threshold at the empirical 5th percentile of $R^{2}$ on the HAM10000
validation split ($\tau=0.9426$ at tap~4), which guarantees that at
least 95\% of genuine in-domain images are labelled in-distribution. We
also report the mean normalised per-feature shift
\[
R_{\mathrm{shift}}
= \frac{1}{D}\sum_{d=1}^{D}
\left|\frac{\mu_{tgt,d}-\mu_{src,d}}{\sigma_{src,d}}\right|,
\]
analogous to the per-layer $R$ values of SIMLea.

Because ChromaSwap tones are calibrated to real skin
(Section~\ref{sec:gap}), this same test applies to the augmented images
themselves. That is what we mean by keeping samples in-distribution or
only mildly out-of-distribution: the augmented data can be audited with
the same instrument as the external data, and an arbitrary texture could
not be.

\subsection{Evaluation Protocol}

Because T0 is perfectly class-balanced, accuracy equals balanced
accuracy throughout. The sets T0, T1 and T\_swap are those defined in
Section~\ref{ssec:data} and Table~\ref{tab:notation}. The evaluation
spans seven axes:

clean accuracy on T0; an overfitting check using train, validation and
test accuracy triples; an analysis of attribution-map stability across
the training set; an attempt to group T0 by skin tone using the
Individual Typology Angle (ITA); external dark-skin generalisation on T1
with $R^{2}$ covariate-shift filtering; background-counterfactual
attention and decision stability; and adversarial robustness under the
Fast Gradient Sign Method (FGSM)~\cite{goodfellow2015}. ITA is a single
number computed from the lightness and blue-yellow axis of a skin patch
in CIELAB colour space, commonly used as an automatic stand-in for
Fitzpatrick type when a clinician has not recorded it. We report that
axis because it did not work: inspection showed ITA was measuring
photographic darkness rather than pigmentation, so we treat it as an
unreliable proxy and report no accuracy broken down by ITA group
(Section~\ref{ssec:ita}). All evaluations are single-pass, with no
test-time augmentation, on identical data.

Both background-swap constructions re-composite the lesion using the
composite XAI mask pipeline of Section~\ref{ssec:mask} rather than
ground-truth segmentations, and the resulting masks are imperfect (mean
IoU\,$\approx$\,0.42; Section~\ref{ssec:xaigt}). These masks are
model-dependent, being derived from the attributions of the
Base-Balanced backbone, and we do not claim otherwise. What makes the
comparison fair is that they are computed once and then frozen, so every
model in every table sees the same composited pixels and whatever error
the masks contain is a constant of the test set rather than something
varying between the models compared. The final $11\times11$ dilation
biases toward over-inclusion, preserving the lesion interior at the cost
of a thin ring of original skin. We therefore read T\_swap as a
\emph{relative} robustness probe, not an absolute clinical accuracy.
\subsection{Code Availability}
The implementation of the CAMEO framework, including the SSIM 
stability screen, ruler artifact suppression, composite mask 
generation, ChromaSwap augmentation pipeline, and the $R^{2}$ 
covariate-shift filter, will be made available through the 
Human-Centred AI Lab at the University of New Brunswick at 
\url{https://github.com/Lab-Human-centered-ai/Cameo_project}. The repository 
will include all scripts required to reproduce the experiments, 
evaluation manifests, and the frozen T0 and T\_swap test sets 
reported in this paper.
% ---------------------------------------------------------------
\section{Results}

Throughout this section, \textbf{T0} is the frozen class-balanced
in-distribution HAM10000 test set ($N=342$), \textbf{T1} is the external
dark-skin ISIC set ($N=100$) and \textbf{T\_swap} is the synthetic
background-counterfactual version of T0 ($N=342$); these are defined in
Table~\ref{tab:notation}. \textbf{CAMEO} is our headline model,
\textbf{Base-Balanced} is the class-balanced raw-image baseline and
\textbf{Base-Skewed} is the imbalanced raw-image baseline; all model
names are defined in Table~\ref{tab:modelglossary}.

\subsection{Clean Performance on T0}

Table~\ref{tab:clean} reports performance on the frozen 342-image T0
test set with bootstrap 95\% confidence intervals (10{,}000 resamples).
Base-Balanced remains the strongest model on clean in-distribution data,
with accuracy 0.848 [0.810, 0.886] and AUC-ROC 0.912 [0.881, 0.941].
CAMEO trails Base-Balanced by only 1.2\,pp in accuracy (0.836 [0.795,
0.874]) while achieving the \emph{highest} AUC-ROC (0.913 [0.880,
0.942]). The two metrics answer different questions. Accuracy asks how
many images fall on the correct side of a fixed decision threshold,
whereas AUC-ROC asks how well the model \emph{orders} the images by
melanoma probability, independently of where that threshold is placed.
CAMEO therefore orders the test set at least as well as the baseline
does, and its small accuracy deficit reflects threshold placement rather
than a loss of discriminative information. A paired McNemar
test~\cite{mcnemar1947} on T0 does \emph{not} find a
significant accuracy difference between Base-Balanced and CAMEO
($p=0.62$). CAMEO-Skewed improves over Base-Skewed by 5.3\,pp, which
confirms that XAI-guided augmentation partially compensates for class
imbalance without fully replacing explicit balancing.

\begin{table*}[!t]
\caption{Clean accuracy and related metrics on the frozen balanced test
set T0 ($N=342$). Because T0 is class-balanced, accuracy equals balanced
accuracy, so a single accuracy column is reported. Bracketed values are
bootstrap 95\% confidence intervals (10{,}000 resamples) for the
adjacent metric.}
\centering
\footnotesize
\begin{tabular*}{\textwidth}{@{\extracolsep{\fill}}lcccc@{}}
\toprule
\textbf{Model} & \textbf{Accuracy} & \textbf{Macro F1}
               & \textbf{AUC-ROC} & \textbf{Cohen's $\kappa$} \\
\midrule
Base-Skewed          & 0.743 {\scriptsize[0.696,\,0.789]} & 0.736 & 0.819 {\scriptsize[0.774,\,0.862]} & 0.485 \\
Base-Balanced        & \textbf{0.848} {\scriptsize[0.810,\,0.886]} & 0.848 & 0.912 {\scriptsize[0.881,\,0.941]} & 0.696 \\
\textbf{CAMEO (ours)} & 0.836 {\scriptsize[0.795,\,0.874]} & 0.836 & \textbf{0.913} {\scriptsize[0.880,\,0.942]} & 0.673 \\
CAMEO-Skewed         & 0.795 {\scriptsize[0.751,\,0.836]} & 0.791 & 0.898 {\scriptsize[0.862,\,0.931]} & 0.591 \\
\bottomrule
\end{tabular*}
\label{tab:clean}
\end{table*}

\subsection{Overfitting Check}

\begin{figure}[!htbp]
\centering
\includegraphics[width=\columnwidth,height=0.21\textheight,keepaspectratio]{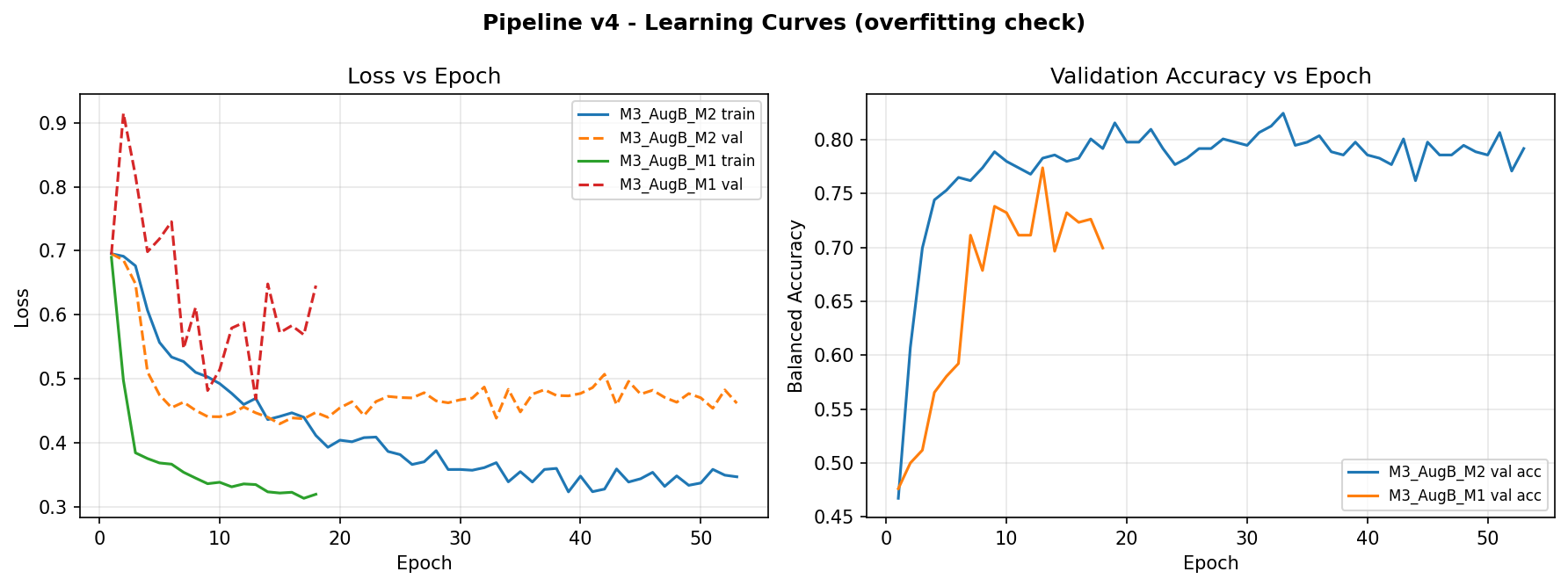}
\caption{Training and validation loss and accuracy against epoch for the
augmented models, showing convergence and the observed gap between
training and validation performance.}
\label{fig:curves}
\end{figure}

Table~\ref{tab:triples} reports the train, validation and test accuracy
triples. The row to read first is \textbf{CAMEO}, the final output model
of this work and the checkpoint carried into every subsequent
experiment; the others are baselines and variants for comparison. CAMEO
shows a train-minus-test gap of 0.159, inside the range spanned by the
baselines and smaller than that of CAMEO-Skewed, so the augmentation
introduces no overfitting problem of its own. Base-Balanced, CAMEO and
CAMEO-Skewed all reach high train accuracy ($\ge 0.98$) with gaps of
0.13 to 0.20, the moderate memorisation typical of fine-tuned CNNs on
small medical datasets, and the largest gap (0.196, CAMEO-Skewed) is
consistent with overfitting to synthetic melanoma backgrounds. The train
estimate for Base-Skewed (0.643) falls \emph{below} its T0 test score
because the 2{,}000-image training subset reflects the heavily
imbalanced distribution under evaluation-only transforms.
Figure~\ref{fig:curves} shows the training curves.

\begin{table}[!htbp]
\caption{Train, validation and test accuracy (overfitting check). Train
accuracy is measured on a random 2{,}000-image subset with evaluation
transforms only. All values are accuracies.}
\centering
\scriptsize
\setlength{\tabcolsep}{3.5pt}
\begin{tabularx}{\columnwidth}{@{}Lcccc@{}}
\toprule
\textbf{Model} & \textbf{Train} & \textbf{Val.}
               & \textbf{Test (T0)} & \textbf{Train $-$ Test} \\
\midrule
Base-Skewed          & 0.643 & 0.735 & 0.743 & $-$0.100 \\
Base-Balanced        & 0.981 & 0.795 & 0.848 & $+$0.133 \\
\textbf{CAMEO (ours)} & 0.995 & 0.824 & 0.836 & $+$0.159 \\
CAMEO-Skewed         & 0.992 & 0.759 & 0.795 & $+$0.196 \\
\bottomrule
\end{tabularx}
\label{tab:triples}
\end{table}

\subsection{XAI Instability Analysis}
\label{ssec:ssim}

Table~\ref{tab:ssim} quantifies explanation stability over all 3{,}000
XAI-extracted training images, measuring how much of the data is safe to
act on. The mean per-image SSIM under input perturbation is 0.758
(median 0.766); 744 images (24.8\%) fall below the 0.7 threshold and are
flagged XAI-unstable, and a further 1{,}559 (52.0\%) exceed the
background-bias criterion, meaning more than 40\% of composite saliency
energy lies in the image border. These two numbers justify the design
choice of Section~\ref{ssec:stability}: on a quarter of the training set
the explanation shifts under imperceptible noise, and on roughly half it
is concentrated where the lesion is not. Repainting backgrounds with no
regard for either fact would propagate the model's own errors into the
dataset, so the pipeline instead routes flagged samples to a stronger
ChromaSwap schedule, keeping the data while applying extra
background-decorrelation pressure where attention is most fragile.

\begin{table}[!htbp]
\caption{XAI-instability statistics over the 3{,}000-image extraction
set. SSIM is the mean structural similarity of the GradCAM map under
three Gaussian input perturbations ($\sigma=0.01$).}
\centering
\scriptsize
\begin{tabularx}{\columnwidth}{@{}Lc@{}}
\toprule
\textbf{Statistic} & \textbf{Value} \\
\midrule
Images analysed                     & 3{,}000 \\
Mean SSIM                           & 0.758 \\
Median SSIM                         & 0.766 \\
SSIM range (min to max)             & 0.451 to 0.967 \\
XAI-unstable (SSIM $<0.7$)          & 744 \,(24.8\%) \\
Background-bias flagged             & 1{,}559 \,(52.0\%) \\
\bottomrule
\end{tabularx}
\label{tab:ssim}
\end{table}

\subsection{ITA Skin-Tone Proxy Validation (Unsuccessful)}
\label{ssec:ita}

We attempted to stratify T0 by skin tone using the Individual Typology
Angle (ITA) computed from non-lesion skin pixels in
CIELAB~\cite{kinyanjui2020}, following prior dermatology-fairness work.
Manual inspection showed that images binned as ``Dark'' by ITA were
frequently dark for reasons \emph{other} than Fitzpatrick skin tone,
including shadows, hair, lesion pigmentation and imaging artefacts. Since
the proxy did not reliably separate skin tones, we do \emph{not} report
stratified accuracy from ITA, relying instead on the external T1 set
(Section~\ref{sec:t1}) and the background-counterfactual probes
(Section~\ref{ssec:swap}).

\subsection{External Dark-Skin Generalisation with Covariate-Shift
Filtering (T1)}
\label{sec:t1}

\subsubsection{Covariate-Shift Quantification}

\begin{figure*}[!htbp]
\centering
\includegraphics[width=\textwidth]{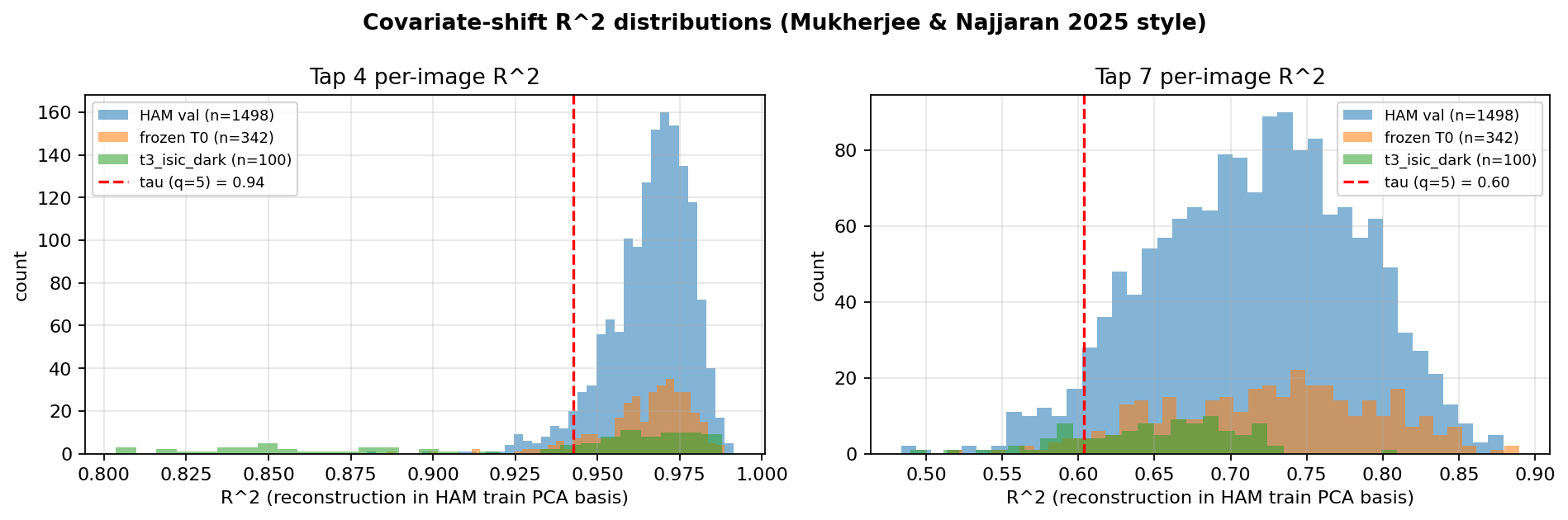}
\caption{$R^{2}$ distributions for the HAM10000 validation split, the
frozen T0 test set and the external T1 dark-skin set. The vertical
dashed line marks the in-distribution threshold $\tau=0.9426$. T0 aligns
with HAM10000 validation, whereas T1 shows a pronounced low-$R^{2}$ tail
that corresponds to out-of-distribution images.}
\label{fig:r2hist}
\end{figure*}

Table~\ref{tab:rshift} reports $R^{2}$ and $R_{\mathrm{shift}}$ for
three sets in the feature space of Base-Balanced. The frozen T0 control
aligns closely with the HAM10000 validation calibration set
($R^{2}=0.965$, in-distribution rate 93.3\%), confirming the filter does
not falsely flag in-domain data. T1 shows mean $R^{2}=0.929$ and
$R_{\mathrm{shift}}=1.751$ at tap~4, about 13 times larger than T0, with
an in-distribution rate of 63\%. More than a third of the dark-skin ISIC
images therefore lie statistically outside the HAM10000-trained feature
manifold, consistent with the known bias of HAM10000 toward
Fitzpatrick~I to III and early-stage flat lesions.
Fig.~\ref{fig:r2hist} shows the distributions and the threshold.

\begin{table}[!htbp]
\caption{Covariate-shift statistics in the Base-Balanced feature space
(tap~4). $R^{2}$ is the per-image reconstruction R-square in the HAM10000
training PCA basis, $R_{\mathrm{shift}}$ is the mean normalised
per-feature shift, and the in-distribution (ID) rate is the fraction of
images with $R^{2}\geq\tau=0.9426$.}
\centering
\scriptsize
\setlength{\tabcolsep}{3.5pt}
\begin{tabularx}{\columnwidth}{@{}Lcccc@{}}
\toprule
\textbf{Set} & $\boldsymbol{N}$ & \textbf{Mean\,$R^{2}$}
             & $\boldsymbol{R_{\mathrm{shift}}}$ & \textbf{ID rate} \\
\midrule
HAM10000 val.\ (calibration) & 1{,}498 & 0.966 & 0.000 & 95.0\% \\
Frozen T0 (control)          &     342 & 0.965 & 0.131 & 93.3\% \\
T1 ISIC dark (external)      &     100 & 0.929 & 1.751 & 63.0\% \\
\bottomrule
\end{tabularx}
\label{tab:rshift}
\end{table}

\subsubsection{Filtered Accuracy on T1}

Table~\ref{tab:filtered} reports the headline external-generalisation
result. On the full 100-image T1 set, Base-Balanced drops from 0.848
accuracy on T0 to 0.640, a loss of 20.8\,pp. Restricting evaluation to
the 63 images flagged in-distribution recovers Base-Balanced to 0.810,
within 3.8\,pp of its T0 score, and recovers CAMEO from 0.660 to 0.794.
These findings isolate two distinct failure modes: genuine
classification error within the trained feature manifold, and covariate
shift beyond that manifold, which cannot fairly be attributed to model
quality. The smaller gain for Base-Skewed is explained by its larger and
noisier training set of 6{,}998 images, which incidentally covers a
broader feature space. After filtering, Base-Skewed (0.857) and
Base-Balanced (0.810) are far closer than their unfiltered gap of
18\,pp.

\begin{table}[!htbp]
\caption{Accuracy on T1 (external dark-skin ISIC set): full set versus
in-distribution (ID) subset ($n_{\text{ID}}=63$, $n_{\text{OOD}}=37$).
$\Delta$ is the accuracy gain obtained by restricting to ID images.}
\centering
\scriptsize
\setlength{\tabcolsep}{3.5pt}
\begin{tabularx}{\columnwidth}{@{}Lcccc@{}}
\toprule
\textbf{Model} & \textbf{Accuracy} & \textbf{Accuracy}
               & $\boldsymbol{\Delta}$ & \textbf{AUC} \\
               & \textbf{(all)} & \textbf{(ID)} & & \textbf{(ID)} \\
\midrule
Base-Skewed          & 0.820 & 0.857 & $+$0.037 & 0.861 \\
Base-Balanced        & 0.640 & 0.810 & $+$0.170 & 0.796 \\
\textbf{CAMEO (ours)} & 0.660 & 0.794 & $+$0.134 & 0.676 \\
CAMEO-Skewed         & 0.630 & 0.778 & $+$0.148 & 0.726 \\
\bottomrule
\end{tabularx}
\label{tab:filtered}
\end{table}

\subsection{Attention and Decision Stability under Background Swap}
\label{ssec:swap}

The clean T0 and T1 numbers show that ChromaSwap affects generalisation,
but not \emph{why}, and the direction of the effect is worth stating
first. ChromaSwap costs 1.2\,pp on T0, which is not statistically
significant, and changes little on the unfiltered T1. What it changes is
behaviour when the background moves: across five seeds it cuts the
accuracy lost between T0 and T\_swap from 8.2\,pp to 2.1\,pp
(Table~\ref{tab:multiseed}), buying stability under tone shift rather
than raw in-distribution accuracy. To test the shortcut-correction
hypothesis behind this, we hold each lesion fixed, swap its background
for several Fitzpatrick-scale synthetic tones, and measure how much the
melanoma decision and the attention of each model change. A model that
ignores the background should be invariant; one exploiting the skin-tone
shortcut should not.

Table~\ref{tab:swap} reports two complementary probes. In the
\emph{decision-stability} probe (150 images by 6 tones), the predicted
label of the baseline Base-Balanced flips on 19.8\% of image-and-tone
pairs, and its melanoma probability spans a mean range of 0.250 across
tones. For CAMEO these fall to 5.4\% and 0.123 respectively, a
$3.6\times$ reduction in label flips and a halving of probability
spread. Wilcoxon signed-rank tests~\cite{wilcoxon1945} on the 150 paired counterfactual
images confirm that both reductions are highly significant (probability
range $p<10^{-13}$; per-image flip count $p<10^{-6}$). In the
\emph{attention-stability} probe (120 images), the SSIM between
GradCAM++ maps computed before and after a dark-background swap rises
from 0.720 for Base-Balanced to 0.842 for CAMEO. \textbf{That is, the
saliency of the augmented model stays on the lesion when the background
changes, whereas the saliency of the baseline does not.}

\begin{figure*}[!t]
\centering
\includegraphics[scale=0.4]{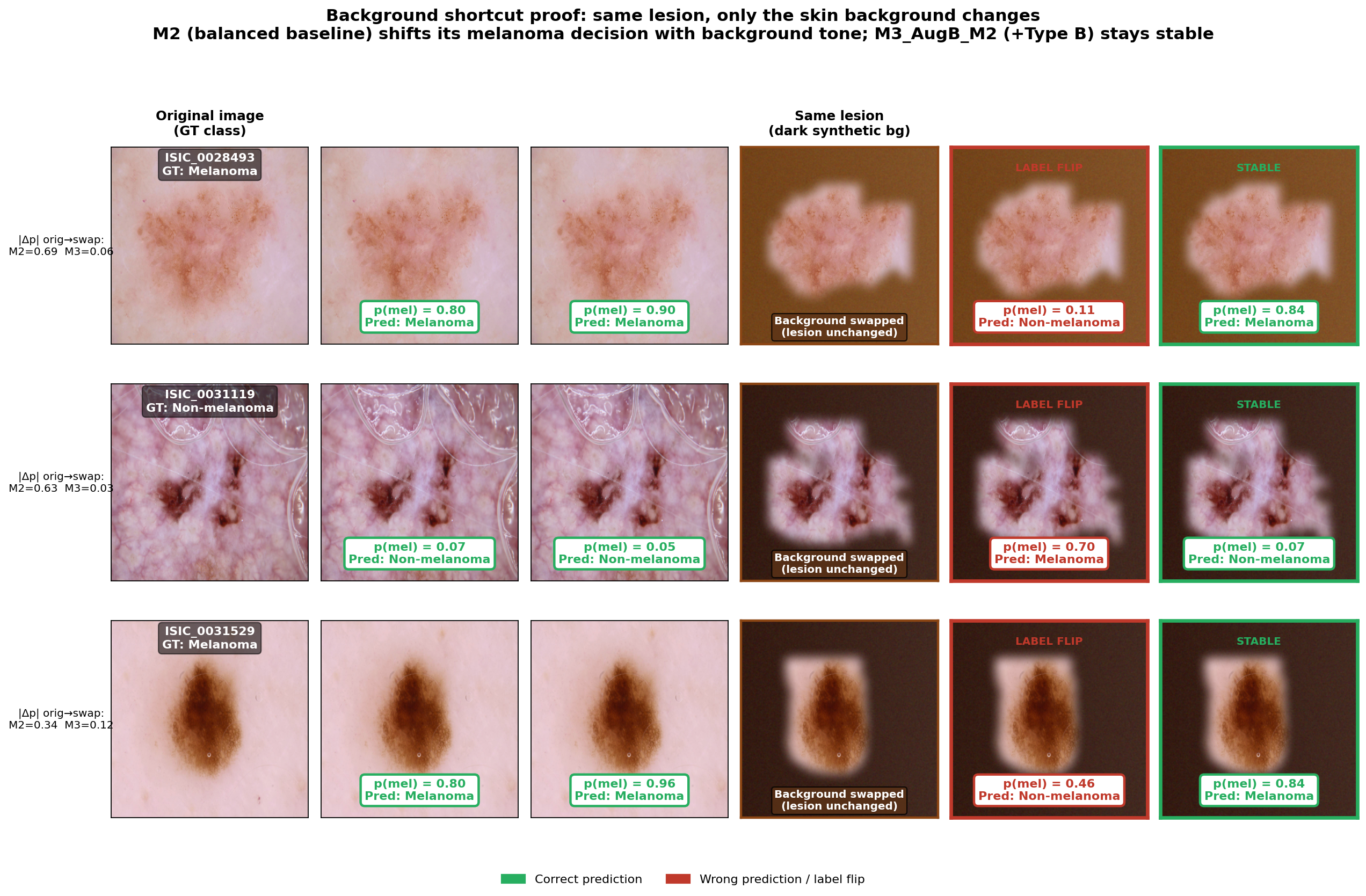}
\caption{Background shortcut proof on three lesions. Columns 1 to 3:
original image with the ground-truth (GT) class and each model's
melanoma probability. Column 4: the \emph{same} lesion with only the
background swapped to a dark synthetic tone. Columns 5 and 6:
predictions after the swap, green for stable or correct and red for a
flip or error. Base-Balanced flips when the background darkens; CAMEO
does not.}
\label{fig:shortcut_proof}
\end{figure*}

\begin{figure*}[!t]
\centering
\includegraphics[scale=0.45]{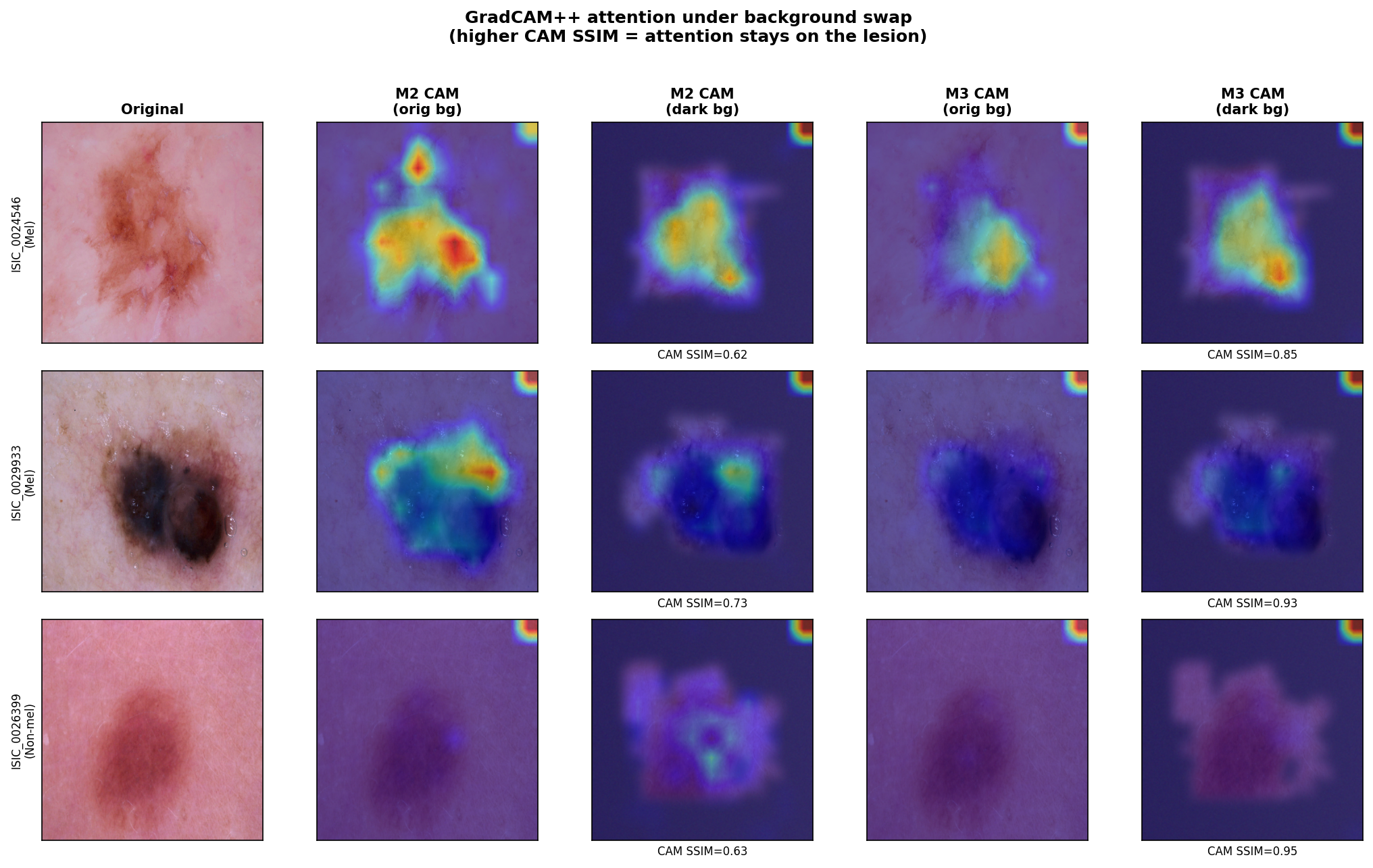}
\caption{GradCAM++ attention on the same lesion with its original and a
dark synthetic background. Base-Balanced (columns 2 and 3) drifts onto
the background when the tone changes, giving a low map SSIM, whereas
CAMEO (columns 4 and 5) stays on the lesion. Per-image SSIM is printed
beneath each pair.}
\label{fig:swap_cam}
\end{figure*}

\begin{figure*}[!t]
\centering
\includegraphics[scale=0.5]{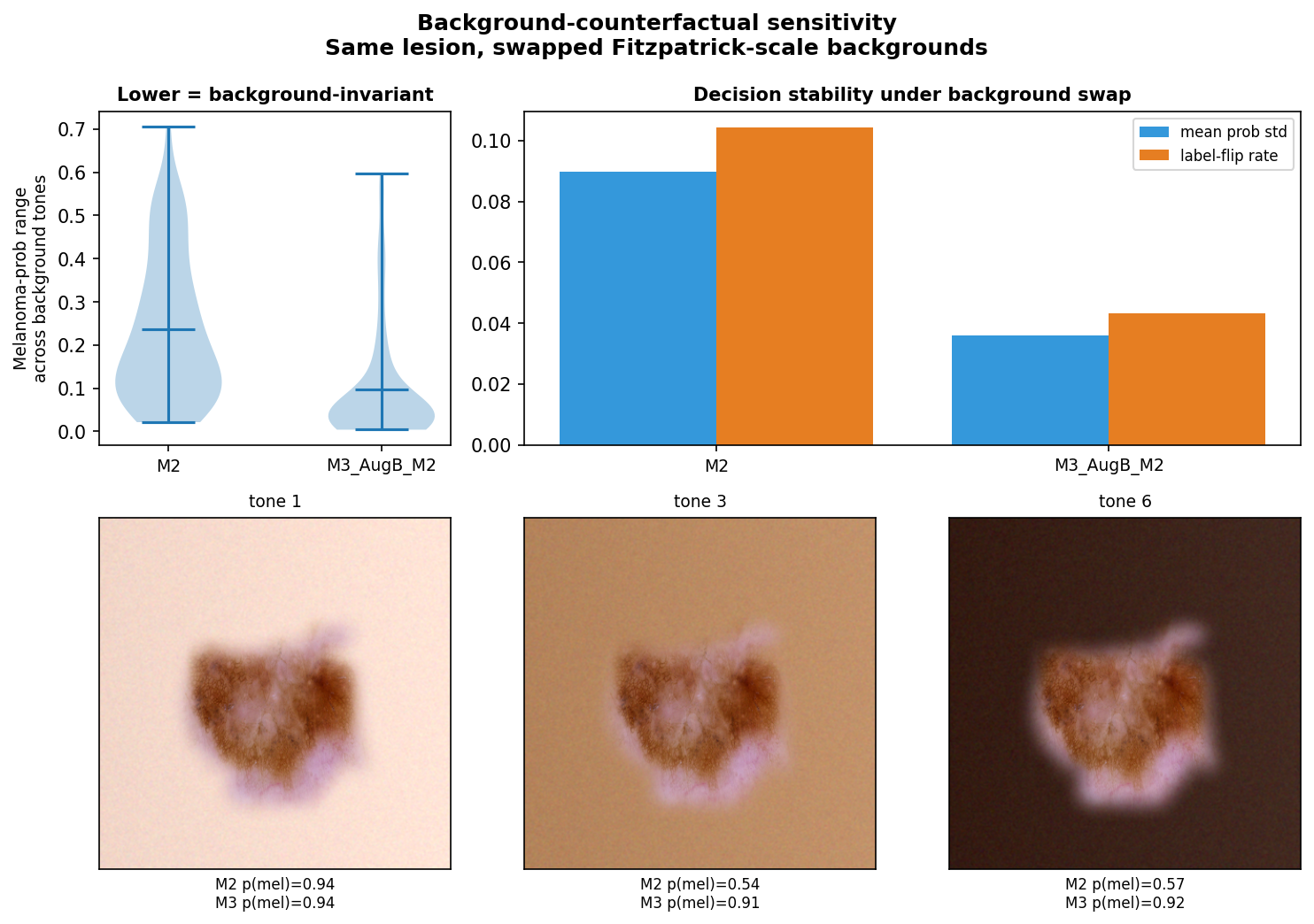}
\caption{Decision stability under background swap. Left: per-image
melanoma-probability range across tones, lower being more invariant.
Right: mean probability standard deviation and label-flip rate. Bottom:
a lesion whose probability under Base-Balanced swings widely with
background tone while CAMEO stays nearly constant.}
\label{fig:swap_prob}
\end{figure*}

Figure~\ref{fig:shortcut_proof} gives the clearest evidence. In the
third case (ISIC\_0031249, ground truth non-melanoma) the label of
Base-Balanced \emph{flips} to melanoma when only the background darkens,
its probability rising from 0.46 to 0.90, whereas CAMEO stays at 0.09.
Figure~\ref{fig:swap_cam} shows the complementary GradCAM++ view, in
which the attention of Base-Balanced scatters onto the surrounding skin
while that of CAMEO stays locked on the lesion, and
Figure~\ref{fig:swap_prob} summarises the population statistics.

We want to point out a caveat before concluding. On \emph{clean}
in-distribution images the mean lesion-attention ratio, the fraction of
GradCAM++ energy inside the ground-truth lesion, is slightly higher for
Base-Balanced (0.55) than for CAMEO (0.44), consistent with its
marginally higher clean accuracy. The benefit of ChromaSwap is therefore
not sharper localisation on easy in-domain images.

What CAMEO delivers instead is \textbf{invariance of the diagnosis to
background skin tone}, and this is the result the paper is built around.
A model whose decision flips on 19.8\% of background swaps will behave
differently on patients who differ only in skin tone; at 5.4\% that
behaviour is largely removed, and the attention maps confirm the
remaining decisions are made on the lesion rather than the skin around
it. Crucially, this is achieved by editing training images that already
exist, requiring no demographic labels, no new dark-skin cohort and no
additional expert annotation, which are precisely the resources every
data-collection route to fairness presupposes.

\begin{table}[!htbp]
\caption{Background-counterfactual robustness. Decision stability: lower
probability range and lower flip rate are better (150 images, 6 tones).
Attention stability: higher GradCAM++ map SSIM before versus after a
background swap is better (120 images).}
\centering
\scriptsize
\setlength{\tabcolsep}{4pt}
\begin{tabularx}{\columnwidth}{@{}Lccc@{}}
\toprule
\textbf{Model} & \textbf{Prob.\ range} & \textbf{Flip rate}
               & \textbf{CAM SSIM} \\
\midrule
Base-Balanced        & 0.250 & 19.8\% & 0.720 \\
\textbf{CAMEO (ours)} & \textbf{0.123} & \textbf{5.4\%} & \textbf{0.842} \\
\bottomrule
\end{tabularx}
\label{tab:swap}
\end{table}

\subsection{XAI Mask Validation against Ground-Truth Lesion Masks}
\label{ssec:xaigt}

Table~\ref{tab:xaigt} compares composite XAI masks with the ISBI 2018
HAM10000 ground-truth lesion segmentations~\cite{tschandl2019seg} on
2{,}332 matched images. A mean Dice of 0.570 (median 0.611) confirms
that the saliency-derived masks substantially overlap the clinical
lesion outline, and the slight drop on melanoma (Dice 0.545 versus
0.576) is consistent with its more diffuse, irregular borders.
Critically, our masks reach mean IoU\,=\,0.422 without any bounding-box
or point prompt, without fine-tuning on ground-truth masks and without
any external segmentation model, against the without-prompt SkinSAM
baseline of IoU\,$\approx$\,0.16~\cite{skinsam2023}. Integrated
Gradients alone is a markedly weaker localiser than the fused signal,
and suppressing the ruler artifact (Section~\ref{ssec:mask}) improves
overlap, so we use the ruler-corrected composite throughout rather than
relying on the failure patterns of any single explanation method.

Table~\ref{tab:unsup} compares our masks against classical unsupervised
baselines, and we report the outcome plainly: global Otsu
thresholding~\cite{otsu1979} achieves \emph{higher} raw IoU than our
composite mask (0.552 versus 0.427, median 0.626 versus 0.447). Otsu is
a genuinely strong overlap segmenter on HAM10000, not a strawman, but
its cost lies elsewhere. It segments by global grayscale intensity and
so assumes the lesion is the darkest region, an assumption that fails on
amelanotic and hypopigmented lesions, on images with dark corners, hair
or vignetting, and in precisely the population this paper is about: on
Fitzpatrick V and VI skin the intensity difference between diseased and
healthy skin is much smaller, for the same reason erythema becomes hard
to see as pigmentation increases~\cite{aoki2026,forsyth2025}. An
intensity-threshold segmenter therefore degrades exactly where a
fairness-oriented pipeline needs it to hold, which makes raw IoU on a
light-skinned benchmark a flattering measure of its suitability.
Consistent with that fragility, the per-image IoU of Otsu is more
dispersed (standard deviation 0.236 versus 0.171) and its
catastrophic-failure rate higher, at 6.8\% of images below IoU 0.1
against 4.1\% for ours. Our masks are instead \emph{model-faithful} and
need no intensity heuristic, no annotation and no external model.

\begin{figure*}[!htbp]
\centering
\includegraphics[scale=0.5]{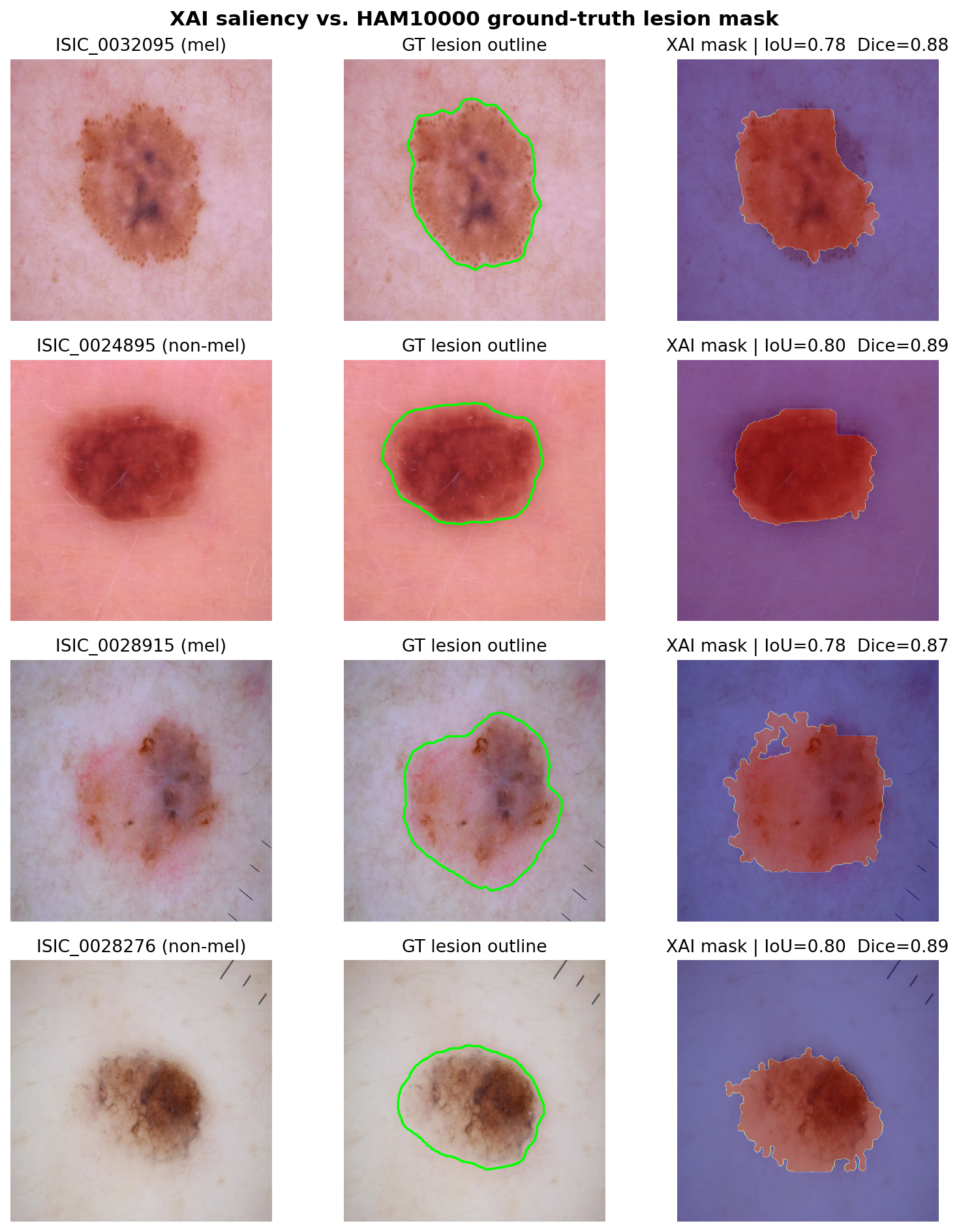}
\caption{Qualitative validation of composite XAI masks against HAM10000
ground-truth lesion segmentations. The green outline is the clinical
lesion mask and the heatmap overlay is the binary XAI mask used for
ChromaSwap augmentation.}
\label{fig:xaigt}
\end{figure*}

We stress that raw mask overlap is not the robustness lever of this
paper. As the random-background control shows (Section~\ref{ssec:tier2}),
background replacement drives robustness largely independently of the
mask source. Whether Otsu-derived masks paired with random backgrounds
match our robustness is a specific experiment that we scope in
Section~\ref{sec:limits}. Figure~\ref{fig:xaigt} shows qualitative
overlays.

\begin{table}[!htbp]
\caption{Composite XAI mask versus the HAM10000 ground-truth lesion mask
(ISBI 2018; Tschandl et al.~\cite{tschandl2019seg}). IoU is
Intersection over Union.}
\centering
\scriptsize
\setlength{\tabcolsep}{3.5pt}
\begin{tabularx}{\columnwidth}{@{}Lcccc@{}}
\toprule
\textbf{Group} & $\boldsymbol{N}$ & \textbf{Mean IoU}
               & \textbf{Mean Dice} & \textbf{Median Dice} \\
\midrule
All          & 2{,}332 & 0.422 & 0.570 & 0.611 \\
Non-melanoma & 1{,}858 & 0.428 & 0.576 & 0.618 \\
Melanoma     &     474 & 0.397 & 0.545 & 0.577 \\
\bottomrule
\end{tabularx}
\label{tab:xaigt}
\end{table}

\begin{table}[!htbp]
\caption{Unsupervised segmentation baselines against ground truth
($N{=}2{,}332$). Otsu achieves higher raw Intersection over Union (IoU)
on this light-skinned benchmark, but it relies on an intensity
assumption that weakens on darker skin and it does not isolate the
model-attributed lesion interior that ChromaSwap augmentation needs.}
\centering
\scriptsize
\setlength{\tabcolsep}{4pt}
\begin{tabularx}{\columnwidth}{@{}Lc@{}}
\toprule
\textbf{Method} & \textbf{Mean IoU} \\
\midrule
Otsu (grayscale)~\cite{otsu1979}             & 0.552 \\
Composite XAI mask (ours)                 & 0.427 \\
HSV colour threshold                      & 0.426 \\
SkinSAM without prompt~\cite{skinsam2023} & $\approx$0.16 \\
\bottomrule
\end{tabularx}
\label{tab:unsup}
\end{table}

\subsection{Exhaustive Fitzpatrick Augmentation Showcase}
\label{ssec:fitz}

\begin{figure*}%[!t]
\centering
\includegraphics[scale=0.45]{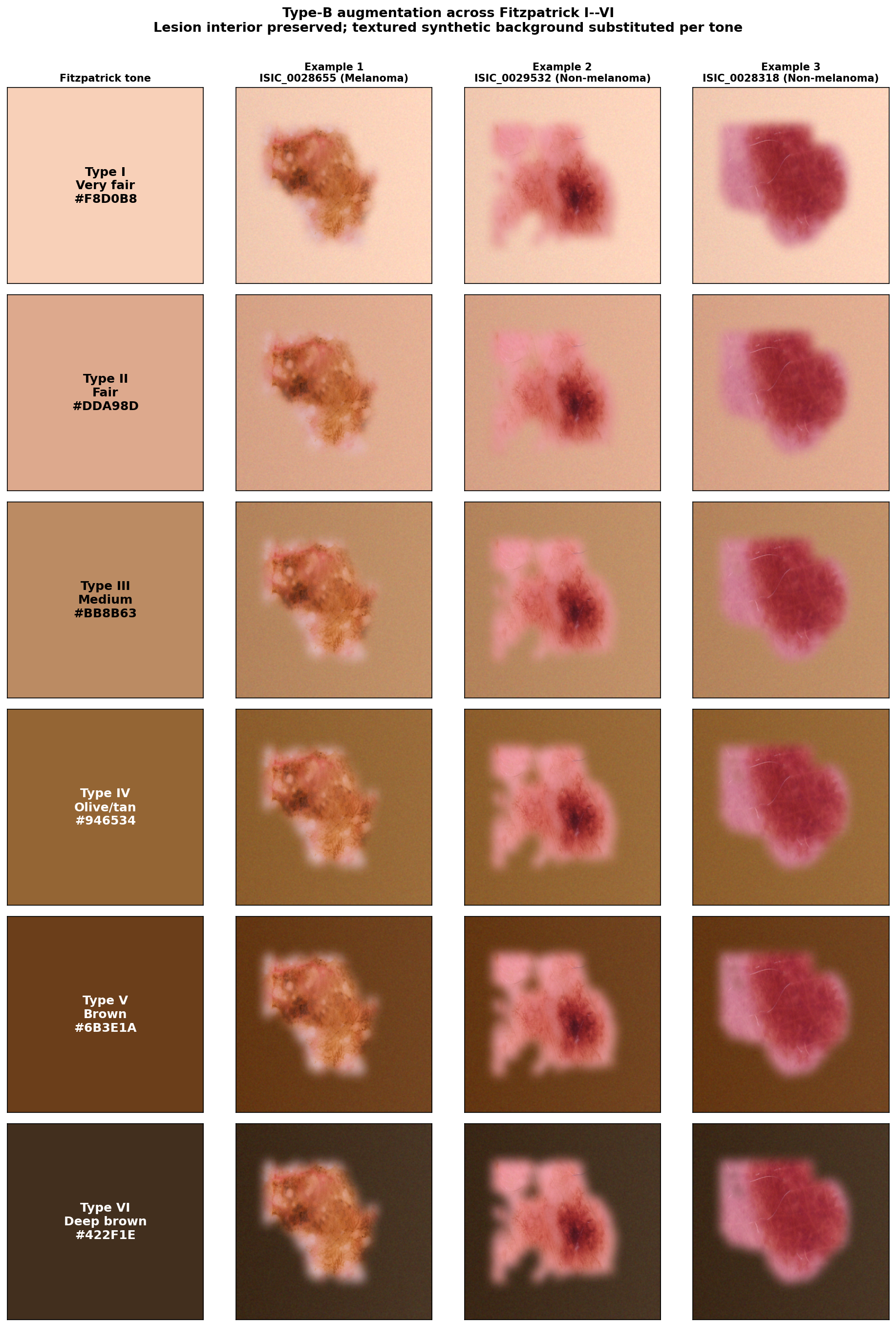}
\caption{Exhaustive ChromaSwap augmentation across Fitzpatrick I to VI
(rows), with three examples each (columns). The same lesions are carried
through every row, so the effect of tone level reads down each column.
The lesion interior is preserved while the background is substituted
with a textured tone. Hex codes and labels are shown at the left.}
\label{fig:fitz}
\end{figure*}

Figure~\ref{fig:fitz} documents ChromaSwap across the full Fitzpatrick
scale (I to VI) with three worked examples per level. The figure shows
how tone replacement behaves across the scale, so the same three lesions
are carried through all six levels rather than re-selected per level.
Mask quality is not a variable here, since the augmentation is applied
identically whatever the mask looks like; the distribution of mask
quality over the whole dataset, including failures, is reported without
selection in Tables~\ref{tab:xaigt} and~\ref{tab:unsup}, where 4.1\% of
images fall below IoU 0.1. In every case the lesion interior is
preserved pixel for pixel while the surrounding skin is replaced with a
textured synthetic tone of the target type. The augmentation therefore
behaves consistently across the whole tone range, including the darkest
and most under-represented types, which is the mechanism by which the
model is exposed to, and desensitised to, skin-tone variation. It is
also what keeps augmented images recognisable as dermoscopic images of
human skin rather than anomalies (Section~\ref{ssec:r2}).

\subsection{Augmentation Ablations}
\label{ssec:ablation}

This subsection isolates which part of the augmentation recipe does the
work. As described in Section~\ref{ssec:data}, the intermediate
checkpoints from pipeline versions~2 and~3 each differ from the final
model in only one or two design decisions, which makes them a useful
ablation. We label them \textbf{Legacy-} to keep them distinct from the
headline v4 model, CAMEO (checkpoint \texttt{m3\_augb\_m2\_v4.pt}), and
evaluate all of them on the frozen T0 and T\_swap manifests without
retraining.

Three augmentation strategies appear in the table, all ours rather than
published methods, so we define them here. \emph{ChromaSwap} is the full
background replacement of Section~\ref{ssec:chromaswap}: the lesion
interior is copied and the surrounding skin repainted with a textured
Fitzpatrick tone. \emph{LesionJitter} is a weaker variant that leaves
the background alone and applies colour and brightness jitter to the
lesion interior only, testing whether localised perturbation of any kind
suffices. \emph{LesionJitter+B}, written ``A $+$ B'' in the table,
applies both. \emph{FlatBG} is the version~2 fill, replacing the
background with a single constant colour instead of a texture, included
to show what happens when the replacement is not made to look like skin.

Reading Table~\ref{tab:augablation}, ChromaSwap background replacement
is the primary driver of robustness: Legacy-ChromaSwap reaches 0.810 on
T\_swap against
0.792 for both LesionJitter and LesionJitter+B, while all three match
within 2\,pp on T0, so editing the lesion is no substitute for editing
the background. Sensitivity to the mask threshold is modest, with
$p_{60}$ to $p_{80}$ spanning 0.789 to 0.810 on T\_swap. Texture, by
contrast, is essential: Legacy-FlatBG collapses to 0.564 on T\_swap
despite reaching 0.787 on T0. That collapse is the clearest evidence
that an unrealistic background does not debias a model but hands it a
new shortcut, and it is why CAMEO uses calibrated skin textures.

The remaining rows compare training recipes. Legacy-ChromaSwap trains on
the full unbalanced HAM10000 with ChromaSwap only, whereas CAMEO uses
the balanced core with symmetric textured ChromaSwap from ImageNet
initialisation. Legacy-Textured (\texttt{m3b\_v3.pt}) warm-starts from
Base-Balanced and matches CAMEO on T0 (both 0.836) but is less robust on
T\_swap (0.787 versus 0.810), consistent with warm-starting from a model
that has already learned the background shortcut.

\begin{table*}[!t]
\caption{Legacy augmentation ablations (v2, v3 and threshold-sweep
checkpoints; these are \emph{not} the headline CAMEO v4 model, see the
text). All checkpoints are evaluated on the frozen T0 and T\_swap
manifests without retraining. T\_swap is the synthetic
background-counterfactual set and not the external T1 set.}
\centering
\scriptsize
\setlength{\tabcolsep}{4pt}
\begin{tabularx}{\textwidth}{@{}s{0.78} s{1.22} c c@{}}
\toprule
\textbf{Checkpoint} & \textbf{Pipeline / recipe}
                    & \textbf{T0 accuracy} & \textbf{T\_swap accuracy} \\
\midrule
Legacy-LesionJitter  & comparison / unbalanced $+$ LesionJitter & 0.798 & 0.792 \\
Legacy-ChromaSwap    & comparison / unbalanced $+$ ChromaSwap   & 0.827 & 0.810 \\
Legacy-Combined      & comparison / unbalanced $+$ A $+$ B      & 0.827 & 0.792 \\
Legacy-ChromaSwap-p70 & threshold sweep                         & 0.825 & 0.807 \\
Legacy-ChromaSwap-p80 & threshold sweep                         & 0.807 & 0.789 \\
Legacy-FlatBG        & v2 flat background                       & 0.787 & 0.564 \\
Legacy-Textured      & v3 warm start from Base-Balanced         & 0.836 & 0.787 \\
\midrule
\textbf{CAMEO (ours, v4)} & balanced core $+$ symmetric ChromaSwap, from scratch & \textbf{0.836} & \textbf{0.810} \\
\bottomrule
\end{tabularx}
\label{tab:augablation}
\end{table*}

\subsection{Multi-Seed Variance and Debiasing Baselines}
\label{ssec:tier2}

% BEGIN TIER2-MULTISEED
\begin{table}[!htbp]
\caption{Multi-seed variance. Mean $\pm$ standard deviation of
accuracy over five seeds on the frozen T0 and T\_swap manifests, where
T\_swap is the synthetic background-counterfactual set of
Table~\ref{tab:notation} and \emph{not} the external T1 set. ``Accuracy
drop'' is the T0 to T\_swap accuracy loss.}
\centering
\footnotesize
\setlength{\tabcolsep}{3.5pt}
\begin{tabularx}{\columnwidth}{@{}Lccc@{}}
\toprule
\textbf{Model} & \textbf{T0 accuracy} & \textbf{T\_swap accuracy}
               & \textbf{Accuracy drop} \\
\midrule
Base-Balanced & $0.827\pm0.016$ & $0.745\pm0.031$ & $0.082\pm0.035$ \\
\textbf{CAMEO (ours)} & $0.836\pm0.008$ & $\mathbf{0.815\pm0.017}$ & $\mathbf{0.021\pm0.022}$ \\
\bottomrule
\end{tabularx}
\label{tab:multiseed}
\end{table}
% END TIER2-MULTISEED

% BEGIN TIER2-DEBIAS
\begin{table*}[!t]
\caption{Head-to-head debiasing baselines, all evaluated on the same
T\_swap manifest. ``CAMEO (ours)'' is the v4 headline checkpoint of
Table~\ref{tab:clean}; the three baselines are single-seed (42) retrains
on the same balanced core with identical hyper-parameters. The last
column is the two-sided McNemar $p$ value against ours.}
\centering
\scriptsize
\setlength{\tabcolsep}{4pt}
\begin{tabularx}{\textwidth}{@{}s{0.74} s{1.26} c c c@{}}
\toprule
\textbf{Model} & \textbf{Method} & \textbf{T0 accuracy}
               & \textbf{T\_swap accuracy} & \textbf{$p$ vs.\ ours} \\
\midrule
\textbf{CAMEO (ours)} & XAI ChromaSwap (Fitzpatrick) & 0.836 & 0.810 & n/a \\
CAMEO-RandomBG & Random non-Fitzpatrick ChromaSwap & 0.830 & 0.816 & 0.87 \\
BiasAdv & BiasAdv-style conflicting background~\cite{biasadv2023} & 0.830 & 0.798 & 0.62 \\
AdvDebias-GRL & Adversarial tone debiasing by gradient reversal~\cite{yang2023} & 0.833 & 0.734 & \textbf{$<$0.01} \\
\bottomrule
\end{tabularx}
\label{tab:debias}
\end{table*}
% END TIER2-DEBIAS

To address single-seed variance and the absence of head-to-head
debiasing comparisons, we retrain Base-Balanced and CAMEO with five
random seeds and implement three baselines on the same balanced core:
(1)~random non-Fitzpatrick background replacement, labelled
CAMEO-RandomBG, which controls whether Fitzpatrick-specific tones are
necessary; (2)~adversarial tone debiasing by gradient reversal on border
luminance~\cite{yang2023}, which uses the gradient-reversal layer of
Ganin et~al.~\cite{ganin2016} to stop the encoder from retaining tone
information; and (3)~BiasAdv-style bias-conflicting ChromaSwap samples
generated against a frozen biased classifier, namely
Base-Skewed~\cite{biasadv2023}. All of these use the T\_swap manifest,
so they are \emph{not} comparable with the external T1 numbers of
Section~\ref{sec:t1}.

\paragraph{Multi-seed robustness.}
Table~\ref{tab:multiseed} shows that the robustness gain is not a
single-seed artifact. Across five seeds, CAMEO improves T\_swap accuracy
over Base-Balanced in \emph{every} seed, with a mean gain of 7.0\,pp and
a range of 1.8 to 13.2\,pp (paired Wilcoxon $p=0.0625$, the minimum
attainable at $n=5$). It reduces the T0 to T\_swap accuracy drop nearly
four-fold ($0.021\pm0.022$ versus $0.082\pm0.035$) while roughly halving
seed-to-seed variance ($\pm0.017$ versus $\pm0.031$), so the augmented
model is both more robust and more \emph{stable}.

\paragraph{Is Fitzpatrick calibration necessary?}
The random-background baseline is the most important control, and
Table~\ref{tab:debias} gives the comparison on the identical 342 T\_swap
images. Our model scores 0.810 and CAMEO-RandomBG scores 0.816. A paired
McNemar test~\cite{mcnemar1947}, the standard test for comparing two
classifiers on the same samples, finds this gap not significant
($p=0.87$; the bootstrap confidence interval of $[-2.9,+4.1]$\,pp spans
zero), and CAMEO-RandomBG lands inside our five-seed spread of
$[0.787,0.833]$. We therefore do \emph{not} claim that
Fitzpatrick-specific tones are necessary for the robustness gain.

Instead, this control \emph{clarifies the mechanism}. The active
ingredient is \textbf{debiasing through lesion-preserving background
decorrelation}, of which Fitzpatrick calibration is a principled,
clinically realistic instance: it matches random backgrounds on
robustness while producing plausible dermoscopic images rather than
out-of-distribution textures, keeping augmented samples near the
distribution the model must serve (Section~\ref{ssec:r2}) and supporting
a fairness reading that arbitrary backgrounds cannot. Against the
established methods it is significantly better than adversarial
gradient-reversal debiasing~\cite{yang2023,ganin2016} (0.810 versus
0.734, $p<0.01$) and tied with BiasAdv~\cite{biasadv2023} (0.810 versus
0.798, $p=0.62$), while remaining annotation-free and interpretable.

\paragraph{What, then, does the XAI machinery contribute?}
If plain background replacement suffices for robustness, one may ask why
the stability screen, the masking and the calibration are needed. The
\emph{robustness} gain does come from background replacement and is
largely agnostic to mask source and palette; the XAI pipeline
contributes three things orthogonal to it. First, the stability screen
\textbf{decides which data is usable at all}: Table~\ref{tab:ssim} shows
a quarter of the extraction set produces explanations that move under
imperceptible noise, and editing images on that basis unknowingly would
inject model-specific artifacts into the training set. No
random-background baseline provides this. Second, it gives
\textbf{annotation-free, model-faithful localisation} needing no
segmentation labels, no foundation model and no intensity assumption of
the kind that makes Otsu fail on darker skin
(Section~\ref{ssec:xaigt}). Third, it yields \textbf{interpretability
and auditability} of what the model attends to.

\subsection{Qualitative Mask and Augmentation Analysis}

\begin{figure*}[!t]
\centering
\includegraphics[scale=0.5]{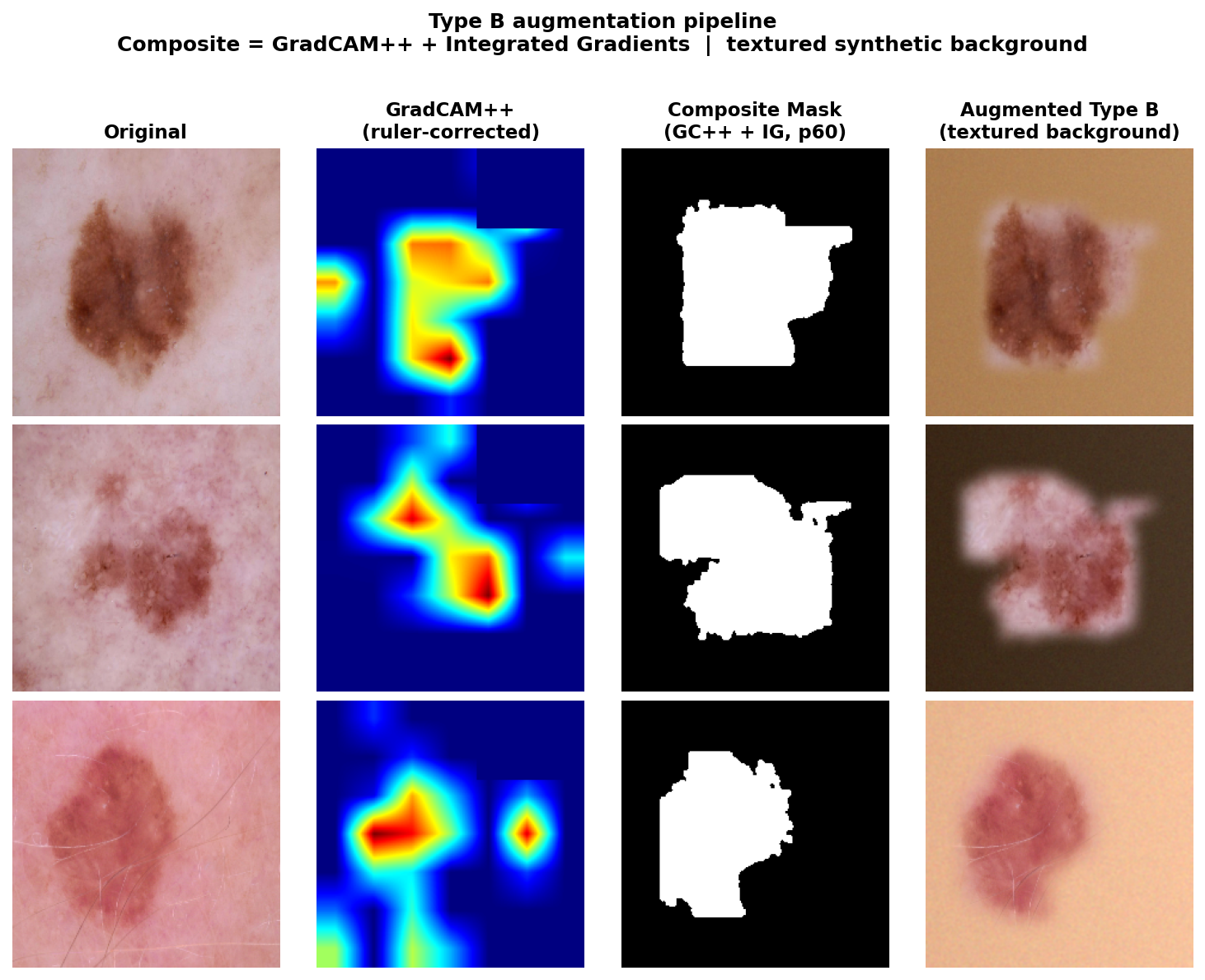}
\caption{Qualitative pipeline results on three representative examples
(rows). From left to right: original image, ruler-corrected GradCAM++
heatmap, cleaned composite mask (GradCAM++ plus Integrated Gradients,
$p_{60}$), and the ChromaSwap augmented image with a Fitzpatrick-scale
background.}
\label{fig:qual}
\end{figure*}

Figure~\ref{fig:qual} illustrates the end-to-end pipeline: the original
image with the upper-right ruler artifact, the ruler-corrected GradCAM++
heatmap, the morphologically cleaned binary mask, and the ChromaSwap
augmented image. The cleaned masks localise the lesion compactly and
exclude the ruler zone, and the augmented images preserve lesion
structure while presenting realistic, diverse skin-tone backgrounds.

\subsection{FGSM Adversarial Robustness}

Table~\ref{tab:fgsm} reports adversarial accuracy on T0 under Fast
Gradient Sign Method (FGSM) perturbations at several epsilon values. A
key finding is that \emph{clean accuracy and adversarial robustness are
orthogonal}. Base-Balanced leads on clean accuracy (0.848) but is the
most brittle model at small epsilon values, dropping to 0.085 at
$\varepsilon=0.01$, while CAMEO-Skewed is the most resilient at low
epsilon (0.272 at $\varepsilon=0.01$) despite its lower clean accuracy.
This brittleness is consistent with our shortcut-learning hypothesis: a
model that relies on high-frequency background features builds sharper
and less robust decision boundaries. ChromaSwap augmentation in CAMEO
improves over Base-Balanced at every epsilon up to 0.05 while retaining
most of the clean accuracy of Base-Balanced.

\begin{table*}[!t]
\caption{FGSM adversarial accuracy on the frozen T0 test set ($N=342$).
Epsilon values are given in normalised input space. All values are
accuracies.}
\centering
\footnotesize
\begin{tabular*}{\textwidth}{@{\extracolsep{\fill}}lcccccc@{}}
\toprule
\textbf{Model} & \textbf{Clean acc.}
  & $\boldsymbol{\varepsilon{=}.001}$
  & $\boldsymbol{\varepsilon{=}.01}$
  & $\boldsymbol{\varepsilon{=}.03}$
  & $\boldsymbol{\varepsilon{=}.05}$
  & $\boldsymbol{\varepsilon{=}.10}$ \\
\midrule
Base-Skewed          & 0.743 & 0.345 & 0.158 & 0.213 & 0.447 & 0.500 \\
Base-Balanced        & 0.848 & 0.213 & 0.085 & 0.105 & 0.377 & 0.500 \\
\textbf{CAMEO (ours)} & 0.836 & 0.325 & 0.193 & 0.219 & 0.415 & 0.500 \\
CAMEO-Skewed         & 0.795 & 0.409 & 0.272 & 0.289 & 0.386 & 0.503 \\
\bottomrule
\end{tabular*}
\label{tab:fgsm}
\end{table*}

% ---------------------------------------------------------------
\section{Discussion}

\textbf{CAMEO} is our headline model (the balanced core
with symmetric ChromaSwap augmentation), \textbf{Base-Balanced} and
\textbf{Base-Skewed} are the two raw-image baselines
(Table~\ref{tab:modelglossary}), \textbf{T0} is the frozen
in-distribution test set, \textbf{T1} is the external dark-skin set and
\textbf{T\_swap} is the synthetic background-counterfactual set
(Table~\ref{tab:notation}).

\textbf{(1) Screening explanations is what makes the rest of the
framework defensible.} Attribution maps carry the biases of the model
that produced them, and Table~\ref{tab:ssim} quantifies it: a
quarter of the training set yields explanations that move under input
noise too small to see, and about half concentrate saliency in the image
border rather than the lesion. A pipeline that skipped this check would,
on a substantial fraction of the data, repaint according to a shortcut
rather than a lesion, writing that shortcut back into the training set
as a new artifact. Because CAMEO measures stability first and converts
the result into a per-sample augmentation strength, low-confidence
explanations become extra debiasing pressure instead of silent
corruption. This is therefore, a crucial step in the pipeline.

\textbf{(2) Class balancing is the dominant in-distribution factor.} The
T0 accuracy of 0.848 for Base-Balanced confirms that training on a
balanced real subset is a powerful, simple baseline; CAMEO trails by
1.2\,pp while matching its AUC-ROC, so ChromaSwap does not harm clean
performance.

\textbf{(3) The fairness consideration requires covariate-shift awareness.} The
apparent failure of Base-Balanced on the full T1 set (0.640) is largely
domain mismatch: 37\% of T1 lies outside the trained feature manifold at
13 times the shift of T0, and restricting to the in-distribution subset
recovers 0.810. A model cannot fairly be penalised for failing on data
statistically foreign to its training distribution.

\textbf{(4) ChromaSwap debiases the classifier by decorrelating the
background.} The counterfactual probes (Section~\ref{ssec:swap}) supply
the mechanism behind the numbers: the decisions and attention of CAMEO
are markedly more invariant to background tone (5.4\% versus 19.8\%
label flips; CAM SSIM 0.84 versus 0.72; both significant at
$p<10^{-6}$), and clean T0 accuracy still does not differ significantly
from Base-Balanced (McNemar $p=0.62$). The controls
(Section~\ref{ssec:tier2}) locate that gain in lesion-preserving
background decorrelation rather than any specific palette, with
Fitzpatrick calibration the clinically realistic instance of the
mechanism.

\textbf{(5) CAMEO-Skewed illustrates the risk of augmentation-only
balancing.} Applying ChromaSwap to the minority class alone causes
overfitting to synthetic melanoma backgrounds (train-minus-test gap
0.196) and lower accuracy on the in-distribution part of T1 (0.778), so
synthetic augmentation \emph{complements} real-data class balancing but
does not \emph{replace} it.

\textbf{(6) Robustness is multidimensional.} The FGSM results show that
clean accuracy, adversarial robustness and distribution-shift robustness
are distinct axes that trade off, so a single accuracy number
understates the practical value of the framework.

% ---------------------------------------------------------------
\section{Limitations and Future Work}
\label{sec:limits}

The names used below are defined in Table~\ref{tab:modelglossary}
(models) and Table~\ref{tab:notation} (datasets: T0 in-distribution
test, T1 external dark-skin test, T\_swap synthetic background swap).

\subsection{Limitations}

Seven limitations bound the claims of this paper.

\begin{itemize}
    \item \textbf{Skin-tone bias of the training data.} HAM10000 is
    heavily skewed toward Fitzpatrick~I to III. We attempted
    stratification by Individual Typology Angle (ITA) on T0 but
    abandoned it after manual inspection showed that images binned as
    ``Dark'' were dark for non-demographic reasons such as shadows, hair
    and lesion pigmentation rather than because the patient had dark
    skin. Our fairness claims therefore rest on the external T1 set and
    on the background-counterfactual probes, not on internal ITA bins.

    \item \textbf{Disease-stage bias of the training data.} HAM10000 is
    predominantly early-stage flat lesions, and this is part of why the
    $R^{2}$ analysis places 37\% of an external dark-skin set outside
    the trained manifold.

    \item \textbf{Scope of the mechanism claim.} Fitzpatrick-specific
    tones are not statistically necessary for the robustness gain, since
    a random background swap performs comparably
    (Section~\ref{ssec:tier2}). We therefore claim Fitzpatrick
    calibration as a principled, fairness-relevant instance of
    background-diversity augmentation that keeps augmented images near
    the deployment distribution, not as the unique cause of the
    improvement. Showing a robustness advantage for tone-specific over
    generic backgrounds would need a larger, genuinely tone-stratified
    cohort.

    \item \textbf{Mask source versus robustness.} Because the gain is
    largely mask-agnostic, one experiment we have not run is
    \emph{Otsu-derived masks paired with random backgrounds}. If that
    ties our pipeline on T\_swap, the XAI masking contributes the
    stability screen, annotation-free faithfulness and interpretability
    rather than the robustness gain itself. It would still need
    repeating on a tone-diverse cohort, since the intensity assumption
    behind Otsu weakens on Fitzpatrick V and VI
    skin~\cite{aoki2026,forsyth2025}.

    \item \textbf{Unstable-sample routing.} XAI-unstable images (24.8\%)
    are routed to a stronger ChromaSwap schedule
    (Section~\ref{ssec:stability}) rather than discarded, preserving
    scarce data and concentrating pressure where attention is most
    fragile. We did not run a controlled discard-versus-reweight
    ablation, so this remains a motivated design choice rather than an
    empirically optimised one; running it is future work.

    \item \textbf{Single-seed baselines.} CAMEO-RandomBG, BiasAdv and
    AdvDebias-GRL are single-seed (42). Replicating each across five
    seeds would allow a full mean-versus-mean comparison, although both
    the head-to-head view and the multi-seed spread already agree that
    CAMEO-RandomBG is not significantly different from ours.

    \item \textbf{Size and construction of the evaluation sets.} T1 is
    small ($N=100$, 37\% out-of-distribution) and distinct from T\_swap;
    the two probe different shifts. Because T\_swap is composited with
    the predicted mask (mean IoU\,$\approx$\,0.42) rather than a
    ground-truth segmentation, its \emph{absolute} accuracies should be
    read with care, since part of any drop may reflect imperfect masking
    rather than background sensitivity. That noise is identical across
    models, so \emph{relative} comparisons remain fair.

    \item \textbf{Narrow experimental scope.} Only one backbone
    (EfficientNet-B4) on one dataset, and the ruler-zone suppression
    parameters (top 30\%, right 40\%) are specific to HAM10000.
\end{itemize}

\subsection{Future Work}

These limitations motivate six directions for future work.

\begin{enumerate}
    \item Train on larger, demographically and stage-stratified cohorts
    such as ISIC 2020 or Diverse Dermatology Images
    (DDI)~\cite{daneshjou2022}.
    \item Apply the SIMLea transfer-feasibility
    checkpoint~\cite{mukherjee2025} to adapt the model to the
    out-of-distribution fraction of T1 using its available labels.
    \item Run the discard-versus-reweight ablation on XAI-unstable
    samples, and the Otsu-mask control described above, to separate the
    contribution of each pipeline component exactly.
    \item Replicate every debiasing baseline across five seeds so that
    the head-to-head comparison becomes mean against mean.
    \item Extend the framework to other conditions whose appearance
    depends on pigmentation. Because CAMEO edits the skin around a
    lesion and leaves the lesion untouched, it transfers directly to
    erythema-driven tasks such as staging pressure injury, grading
    atopic dermatitis and detecting surgical site infection, where the
    diagnostic sign is exactly the one hardest to see on Fitzpatrick V
    and VI skin~\cite{forsyth2025,dhoonmoon2021}.
    \item Evaluate on a second imaging domain such as chest X-ray or
    retinal fundus imaging, to demonstrate the generality of the
    XAI-guided shortcut-correction principle.
\end{enumerate}

% ---------------------------------------------------------------
\section{Conclusion}

This paper presented \textbf{CAMEO}, the Class-Activation-Mapped 
Equitable Overlay framework, which turns a classifier's own 
explanations into an active training signal for robust and fair 
dermoscopic classification. The framework operates in four steps: 
screening every attribution map for stability before allowing it 
to influence training; suppressing a recurring measurement-ruler 
artifact and fusing corrected GradCAM++ and Integrated Gradients 
maps into an annotation-free binary lesion mask; applying 
ChromaSwap, which preserves the lesion pixel-for-pixel while 
replacing the surrounding skin with a textured Fitzpatrick-scale 
synthetic tone; and separating genuine classification error from 
domain mismatch on external dark-skin data using a percentile-
thresholded $R^{2}$ feature-space filter.

CAMEO matches the strong class-balanced baseline within 1.2 
percentage points of clean accuracy on the in-distribution test 
set, a difference that is not statistically significant (McNemar 
$p=0.62$), and achieves the highest AUC-ROC among all models. 
On the external dark-skin set, it recovers to 0.794 accuracy on 
the in-distribution subset after $R^{2}$ filtering, comparable 
to or better than the raw-image baselines on the same fair 
subset. The rate at which the melanoma decision flips when only 
the background skin tone changes falls from 19.8\% to 5.4\%, 
and model attention remains stable on the lesion under background 
swap (GradCAM++ map SSIM 0.84 versus 0.72), results confirmed 
across five random seeds and against established debiasing 
methods including BiasAdv and adversarial gradient-reversal 
debiasing. These gains are achieved without demographic labels, 
without external segmentation supervision, and without any 
additional annotation, by editing training data that already 
exists.

The framework is demonstrated on melanoma classification because 
that task has public labelled data and published fairness 
baselines, but the mechanism edits only the skin surrounding a 
lesion, leaving the lesion intact. It therefore applies directly 
to other conditions whose visual signs are harder to detect on 
darker skin, including pressure injury, atopic dermatitis and 
wound infection, and to other imaging domains where a 
diagnostically irrelevant background drives model shortcuts. 

% ---------------------------------------------------------------
\section*{Funding}

The project was funded by Talent Recruitment Fund from the New Brunswick Innovation Foundation and Harrison McCain Foundation Young Scholars Award, 2024.

% ---------------------------------------------------------------

\end{document}